\documentclass[a4paper,11pt]{article}
\pdfoutput=1 

\usepackage{jheppub} 
\usepackage{gensymb}

\usepackage[T1]{fontenc} 

\def\({\left(} 
\def\){\right)}

\newcommand\rmd { {\rm d} }

\newcommand{\smallz}{{\scriptscriptstyle Z}} %
\newcommand{\mz}{M_\smallz}
\newcommand{\smallw}{{\scriptscriptstyle W}}
\newcommand{\mw}{M_\smallw} 
\newcommand{\gw}{\Gamma_{\smallw}} 
\newcommand{\gz}{\Gamma_{\smallz}}

\def\eps{\epsilon}

\title{\boldmath Higgs boson decay into four leptons at NLOPS electroweak accuracy}

\author[a]{Stefano Boselli,}
\author[b]{Carlo M. Carloni Calame,}
\author[a]{Guido Montagna,}
\author[c]{Oreste Nicrosini}
\author[c]{and Fulvio Piccinini}

\affiliation[a]{Dipartimento di Fisica, Universit\`a di Pavia, and INFN, Sezione di Pavia, 
Via A. Bassi 6, 27100 Pavia, Italy}
\affiliation[b]{Dipartimento di Fisica, Universit\`a di Pavia, Via A. Bassi 6, 
27100 Pavia, Italy}
\affiliation[c]{INFN, Sezione di Pavia, Via A. Bassi 6, 27100 Pavia, Italy}

\emailAdd{stefano.boselli@pv.infn.it}
\emailAdd{carlo.carloni.calame@pv.infn.it}
\emailAdd{guido.montagna@pv.infn.it}
\emailAdd{oreste.nicrosini@pv.infn.it}
\emailAdd{fulvio.piccinini@pv.infn.it}

\abstract{In view of precision studies of the Higgs sector at 
the Run~II 
of the LHC, the improvement of the accuracy of the theoretical prediction is 
becoming 
a pressing issue. 
In this framework, we detail a calculation of the 
full Next-to-Leading Order (NLO) electroweak corrections to Higgs boson decay 
into four charged leptons, by considering the gold-plated channel 
$H \to Z^{(*)} Z^{(*)} \to 2 \ell 2\ell^\prime$, $\ell, \ell^\prime = e, \mu$. We match the
NLO corrections with a QED Parton Shower (PS), in order to simulate exclusive 
multiple 
photon emission and provide novel results at NLOPS 
electroweak accuracy. We compare our NLO predictions to those of the program 
{\tt Prophecy4f} and
present NLOPS phenomenological results relevant for Higgs physics studies, 
with particular 
attention to precision measurements of the Higgs boson mass, spin-parity 
assignment 
and tests of the Standard Model.
Our calculation is implemented in a new code, {\tt Hto4l}, which can be easily 
interfaced to any generator describing Higgs boson production. As an example, 
we provide illustrative results for Higgs production and decay in the 
process $gg \to H \to 4 \ell$ using {\tt POWHEG} with NLOPS accuracy in the
production mode.}

\begin{document} 
\maketitle
\flushbottom

\section{Introduction}
\label{sec:intro}

With the announcement in 2012 of the discovery of a new particle in the 
search for the 
Standard Model (SM) Higgs boson by the ATLAS~\cite{Aad:2012tfa} and 
CMS~\cite{Chatrchyan:2012ufa}  collaborations at the CERN LHC, particle 
physics entered 
a new era. The data collected at the centre-of-mass (c.m.) energies of 7 and 8 TeV 
have been 
analyzed by the two experiments in order 
to establish whether the newly discovered particle is actually the 
boson predicted in the SM as relic of the mechanism of electroweak 
symmetry breaking 
(EWSB)~\cite{PhysRevLett.13.321,Higgs1964132,PhysRevLett.13.508,PhysRevLett.13.585,
PhysRev.145.1156,PhysRev.155.1554}. 

The mass of the observed particle has been precisely measured by studying the 
two cleanest decay channels given by the 
decays into a photon pair and into four charged leptons. The combination of 
the two channels 
$H \to \gamma\gamma$ and $H \to 4\ell~(4e, 4\mu, 2e2\mu)$, which have 
excellent mass 
resolution and 
where excesses with large significance 
are observed \cite{Aad:2014eva,Aad:2014eha,Aad:2014aba,Khachatryan:2014jba,Khachatryan:2014ira,Chatrchyan:2013mxa}, 
presently provides a mass measurement 
of approximately 125~GeV for each experiment, 
with a relative uncertainty of better than
0.2\% for the combined ATLAS-CMS measurement.

Concerning the main production mechanisms of the SM Higgs boson at hadron 
colliders, {\it i.e.} gluon-gluon fusion, vector boson fusion (VBF), 
associated production with a massive vector boson and associated production with top 
quarks, the studies performed at the LHC, based on 
the analysis of individual production signal strengths 
for various decay modes, have provided a 
clear observation of Higgs production through gluon fusion 
and an evidence for VBF production, 
with a significance above the $3\sigma$ level, and for 
associated $VH (V = W,Z)$ production at about $3\sigma$~\cite{Khachatryan:2014jba,Aad:2013wqa}. 

Various tests of the couplings of the new particle to bosons and fermions 
have been 
carried out both by ATLAS and CMS collaborations. 
In particular, the measured ratio of the couplings of the Higgs particle to 
$W$ and 
$Z$ bosons, which is an important probe of the 
EWSB mechanism as fixed by the custodial symmetry, is compatible 
with the SM 
expectation and, more generally, no significant 
deviation from the SM is observed from the coupling strength studies~\cite{Aad:2014eva,Khachatryan:2014jba,Aad:2013wqa}. 
Noticeably, evidence for the direct coupling of the Higgs boson 
to down-type fermions has been reported through the study of the challenging 
decay modes 
Higgs into bottom quarks and $\tau$ leptons~\cite{Aad:2015vsa,Chatrchyan:2014vua}.

Last but not least, the spin and parity quantum numbers of the discovered 
particle 
have been assessed, by means of a systematic analysis 
of its production and decay processes. The data strongly favor the scalar 
nature 
$J^P = 0^+$ of the observed particle, while rejecting other 
non-standard hypotheses ($J^P = 0^-, 1^{\pm}, 2^+$) 
or possibility of $CP$ mixtures at high confidence~\cite{Khachatryan:2014kca,Chatrchyan:2013mxa,Aad:2013xqa}.

All these measurements marked the start of a new era of precision Higgs 
physics and were accompanied by an impressive theoretical effort summarized in three CERN 
reports by the LHC Higgs Cross Section Working
Group~\cite{Dittmaier:2011ti,Dittmaier:2012vm,Heinemeyer:2013tqa}. These studies, as well as the related theoretical
work, are in continuous progress and will continue during the Run II of the 
LHC at higher energies and luminosity.

In this paper, we focus on the Higgs boson decay into four charged leptons, 
{\it i.e.} $H \to Z^{(*)}Z^{(*)} \to 4\ell$, in order to provide novel 
precision predictions of interest for future studies of the Higgs sector at 
the LHC. This decay channel plays a 
particularly relevant r\^ole, as it provides the cleanest experimental 
signature, given by a peak in the four lepton mass spectrum on top of 
a flat and small background. Actually, the $H \to 4\ell$ decay mode allows 
to derive a precise mass measurement in the 
different combinations of lepton final states, to assess the spin-parity 
quantum numbers using sensitive 
angular distributions and to perform 
precision tests of the SM at the level of differential cross sections~\cite{Aad:2014tca}.  
In the off-shell region, the $H \to 4\ell$ data can also be used 
to put constraints on the total width of 
the Higgs boson~\cite{Aad:2015xua, Khachatryan:2014iha}.

In the light of the above motivations, we compute the full set of 
next-to-leading 
order (NLO) electroweak corrections to $H \to Z^{(*)}Z^{(*)} \to 4\ell$, with 
$4 \ell = 4e, 4\mu, 2e2\mu$. We match the NLO corrections to a QED 
Parton Shower (PS), 
in order to simulate multiple 
photon emission exclusively and provide final results at NLOPS electroweak 
accuracy. 
The calculation is available in an
event generator, {\tt Hto4l}\footnote{The 
reference web page is {\tt http:/www.pv.infn.it/hepcomplex/hto4l.html}}, which can be interfaced to any code describing Higgs 
boson 
production. The PS approach is 
based on the ideas first presented in Ref.~\cite{CarloniCalame:2000pz} 
for the simulation of the Bhabha scattering 
process at GeV-scale $e^+ e^-$ 
colliders and later applied to the study of single $W/Z$ 
production in hadronic collisions~\cite{CarloniCalame:2003ux,CarloniCalame:2005vc}. 
The matching procedure 
is a generalization of the method developed in 
Refs.~\cite{Balossini:2006wc,Balossini:2008xr} for the precision calculation 
of $2 \to 2$ 
processes in QED (as encoded in the program 
{\tt BabaYaga@NLO}~\cite{Actis:2010gg,CarloniCalame:2003yt}) 
and also implemented in the event generator {\tt Horace} for the calculation 
of single $W/Z$ hadroproduction processes 
at NLOPS electroweak accuracy~\cite{CarloniCalame:2006zq,CarloniCalame:2007cd}.

The NLO electroweak and QCD corrections to $H \to 4$~fermions decay
processes have been calculated 
in Refs.~\cite{Bredenstein:2006rh,Bredenstein:2006ha} 
and are available in the Monte Carlo (MC) program 
{\tt Prophecy4f}~\cite{Bredenstein:2006nk,Bredenstein:2007ec}, which is used 
in the context of Higgs studies at the LHC for the precision calculation of 
the branching ratios of the decays $H \to Z^{(*)}Z^{(*)}/W^{(*)}W^{(*)} \to 4$~fermions.
 In {\tt Prophecy4f} higher-order photonic corrections are taken into 
account in terms of QED collinear Structure Functions. 
A preliminary study of the impact of the gauge-invariant NLO 
QED and PS corrections to the determination of the Higgs boson mass in 
the $H \to 4\ell$ decay was performed in 
Refs.~\cite{Piccinini:2005iu,CarloniCalame:2006vr}.

The article is organized as follows. In Section \ref{sec:theory} we describe 
the details of our calculation, with 
particular emphasis on the method used for the matching of the NLO electroweak 
corrections with the QED PS. 
In Section \ref{sec:pheno} we present our phenomenological results: 
in Section \ref{sec:nlo} we show a sample of comparisons 
between our predictions and those of {\tt Prophecy4f} as a benchmark of the NLO 
computation, in Section \ref{sec:nlops} we provide results for various observables
at NLOPS EW accuracy, while in Section \ref{sec:pwhg}
we present the results for Higgs production and decay in the 
channel $gg \to H \to 4\ell$ obtained in terms of {\tt POWHEG}~\cite{Nason:2004rx,Frixione:2007vw}
interfaced
to {\tt Hto4l}.
In Section~\ref{sec:conc} we draw our conclusions.

\section{Details of the calculation}
\label{sec:theory}

\subsection{Next-to-leading order (NLO) electroweak corrections}
\label{sec:nloew}

The NLO electroweak corrections to the Higgs boson decay into four charged 
leptons consist 
of QED and purely weak contributions. Since the $H \to 4\ell$ decay is a 
neutral-current 
process, the two subsets are separately gauge invariant and can be computed separately 
as well.

The ${\cal O} (\alpha)$ QED corrections are obtained by attaching a virtual or 
real photon to each charged 
lepton leg. They are expected a priori to provide the dominant contribution, 
as photons which are 
emitted collinear to a lepton give rise to large logarithmic corrections of 
the form $\alpha \log\left( m^2_\ell/Q^2\right)$, 
where $m_\ell$ is the lepton mass and $Q$ some typical energy scale.

The QED virtual corrections comprise vertex and pentagon diagrams 
(in the on-shell renormalization scheme), 
while real photon corrections are induced by the bremsstrahlung process 
$H \to 4\ell + \gamma$. 
The two contributions are separately infrared (IR) divergent but their sum 
is IR-finite. 
We treat the 
IR singularity according to the standard QED procedure of assigning a small 
fictitious 
mass $\lambda$ to the 
photon in the computation of the virtual and real contributions. 
More precisely, the 
Higgs decay width 
associated to the bremsstrahlung correction is separated in two pieces and 
calculated 
as follows 
(in a shorthand notation)
\begin{eqnarray}
\int {\rm d} \Gamma_{\rm real} & = &
\frac{1}{2 M_H} \int_{\lambda}^{\epsilon} {\rm d} \Phi_5 \, |{\cal {M}}_0 
(H \to 4\ell)|^2 
\times ( {\rm eikonal \, \, factor} ) \nonumber\\
&+& \frac{1}{2 M_H}  \int_{\epsilon}^{E_{\rm max}}  {\rm d} \Phi_5 \, |{\cal {M}} (H \to 4\ell + \gamma) |^2
\label{eq:hg}
\end{eqnarray}
where $M_H$ is the Higgs mass, $\epsilon$ is a soft-hard energy separator 
($\epsilon \ll M_H$), 
${\cal {M}}_0 (H \to 4\ell)$ is the 
amplitude of the lowest-order (LO) process $H \to 4\ell$
and
${\cal {M}} (H \to 4\ell + \gamma)$ is the matrix element of the radiative 
decay process 
$H \to 4\ell + \gamma$, 
${\rm d} \Phi_5$ being the 4 leptons plus 1 photon phase space element. 
In Eq.~(\ref{eq:hg}) eikonal factor
stands for the analytical expression of the real radiation correction in 
the soft 
limit $E_\gamma \to 0$. The integral in the first line can be done
analytically (see {\it e.g.}~\cite{'tHooft:1978xw}) and the one in the second line
is performed using standard MC 
integration 
with importance sampling.

The QED virtual counterpart is computed according to the following formula
\begin{eqnarray}
{\rm d} \Gamma_{\rm virt.}^{\rm QED} = \frac{1}{2 M_H} 
{\rm d} \Phi_4
\, 2 \, {\rm Re} \left\{ {\cal {M}}_1^{\rm QED} ({\cal {M}}_0)^* \right\}
\label{eq:hgv}
\end{eqnarray}
where ${\cal {M}}_1^{\rm QED}$ is the one-loop amplitude associated to 
the ${\cal O} (\alpha)$ 
vertex and pentagon diagrams.

We perform the IR cancellation by taking the sum of Eq.~(\ref{eq:hg}) and 
Eq.~(\ref{eq:hgv}) in the numerical
limit $\lambda \to 0$. As a cross-check of the 
calculation, 
we tested that the 
inclusive NLO QED correction coincides with $2 \cdot 3/4\, (\alpha/\pi)$,
which is correctly twice
the inclusive final-state ${\cal O} (\alpha)$ electromagnetic correction to the $Z \to \ell^+ \ell^-$ 
decay \cite{Montagna:1993mf}. 
An important comment is in order here. The tree-level amplitude, as well as 
the amplitude for the real radiation process, contains poles in the phase 
space, corresponding to the points where the momenta of the $\ell^+ \ell^-$ 
pairs and of the $\left(\ell^+ \ell^- \gamma \right)$ system 
cross the zero of the inverse propagators: 
$(p_{\ell^+} + p_{\ell^-})^2 = M_Z^2$ or 
$(p_{\ell^+} + p_{\ell^-} + p_\gamma)^2 = M_Z^2$. These poles are avoided 
considering that the $Z$ boson is an unstable particle, {\em i.e.} its 
propagator contains the finite $Z$-width. This, however, 
would spoil the IR cancellation 
between real and virtual corrections of Eq.~(\ref{eq:hg}) and Eq.~(\ref{eq:hgv}), respectively, 
unless in Eq.~(\ref{eq:hgv}) the QED virtual corrections are calculated 
with unstable $Z$ bosons. The scheme which we adopt 
for the introduction of the width in the $Z$ boson propagator, 
without introducing gauge invariance violations, is the complex mass 
scheme~\cite{Denner:2005fg,Denner:2006ic}, which also allows us 
to include weak loop corrections consistently.\footnote{Actually, the complex mass scheme
is used in our calculation of the weak contributions due to the exchange of 
$W$ bosons and top quarks as well, where a complex top mass is introduced,
in particular, to evaluate loop diagrams with internal top quarks when the Higgs mass 
is close to the ${\rm t} {\rm {\bar t}}$ threshold.}

Concerning the basic features underlying the computation of the complete 
${\cal O}(\alpha)$ virtual corrections, we briefly describe the most important aspects in the following.
Since we work in the 't Hooft-Feynman gauge, 
all the particles present in the spectrum of the SM, 
including the Fadeev-Popov and Higgs-Kibble ghosts, 
are involved in the calculation. The corresponding Feynman diagrams 
include, in addition to two-point functions, 
rank-two tensor three-, four- and five-point functions. 
The related ultraviolet divergencies 
are regularized by means of dimensional regularization. 
The reduction of the tensor $n$-point functions is carried out by means of 
the symbolic manipulation program {\tt FORM}~\cite{Vermaseren:2000nd}. 
The necessary scalar form factors with complex masses 
are evaluated using {\tt Looptools $\!\!$v2.10}~\cite{Hahn:1998yk,Hahn:2000jm}, which implements 
the evaluation of the reduction of tensor five-point integrals 
according to Refs.~\cite{Denner:2002ii,Denner:2010tr}, as well as according 
to Passarino-Veltman reduction techniques~\cite{Passarino:1978jh}. 
The form factors are calculated with complex masses and real 
external squared momenta. 
This is sufficient for the implementation of the ``simplified version 
of the complex renormalization'', as described 
in Refs.~\cite{Denner:2005fg,Denner:2006ic}. The complete expressions 
for the counterterms in the on-shell scheme and for the basic 
self-energy diagrams are taken from Ref.~\cite{Denner:1991kt}. 
Since the collinear singularities associated to 
the photon becoming collinear with one of the leptons are regulated by the 
finite lepton mass, the 
kinematics of the radiative process is calculated including 
exactly the contribution of lepton masses. In order to 
allow the cancellation of soft IR singularities, also 
the tree-level kinematics is calculated with complete lepton 
mass effects taken into account. In addition, this gives 
automatically the correct phase space integration boundaries for the 
diagrams of the virtual contribution where a virtual photon 
is connected to one external lepton 
pair. Although the kinematics is treated exactly, the 
non-IR ${\cal O}(\alpha)$ virtual amplitudes are calculated in the 
approximation of neglecting finite fermion mass effects 
(with the exception of the quark Yukawa couplings, {\it e.g.} in the fermion-loop Higgs
vertex corrections). These contributions are neglected in our calculation as 
they are irrelevant 
in view of a target theoretical accuracy of the order of $0.1$\% and 
their inclusion would make the numerical computation more time consuming. 

In formulae, the Higgs width including one-loop weak corrections is obtained as
\begin{eqnarray}
{\rm d} \Gamma_{\rm virt.}^{\rm weak} = \frac{1}{2 M_H} 
{\rm d} \Phi_4
\, 2 \, {\rm Re} \left\{ {\cal {M}}_1^{\rm weak} ({\cal {M}}_0)^* \right\}
\label{eq:hw}
\end{eqnarray}
where ${\cal {M}}_1^{\rm weak}$ is the one-loop amplitude associated to the 
full set of 
${\cal O} (\alpha)$ weak diagrams.

To check some relevant ingredients of our calculation of one-loop weak 
corrections, we 
compared our predictions for the Higgs decays $H \to ZZ, \gamma\gamma$ at NLO 
electroweak accuracy with those of Ref.~\cite{Kniehl:1993ay}, finding perfect 
agreement.

In conclusion, our predictions for the Higgs boson decay into four leptons 
at NLO EW
accuracy are given by the sum of Eq.~(\ref{eq:hg}), Eq.~(\ref{eq:hgv}) 
and Eq.~(\ref{eq:hw}), 
supplemented with the necessary renormalization conditions.

\subsection{Matching NLO electroweak corrections to QED Parton Shower}
\label{sec:mnlops}

In the present section, we sketch our scheme for the matching of the
NLO EW corrections with a QED PS. We closely follow the
approach already presented and successfully applied to QED processes
at low energies and Drell-Yan $W/Z$ production at hadron
colliders~\cite{Balossini:2006wc,Balossini:2008xr,CarloniCalame:2006zq,CarloniCalame:2007cd}.

On general grounds, the partial decay width corrected for the emission
of an arbitrary number of photons in a PS framework can be written as
follows:
\begin{equation}
  \rmd\Gamma_{\infty}^{PS} = \frac{1}{2 M_H}\Pi(\{p\},\epsilon)\sum_{n=0}^\infty
  \frac{1}{n!}\;\left|{\cal M}_n^{PS}(\{p\},\{k\})\right|^2\;\rmd\Phi_n(\{p\},\{k\})
  \label{eq:dGinftyPS}
\end{equation}
where $\{p,k\}$ stands for the set of the final state lepton and
photon momenta $p_1$, $p_2$,
$p_3$, $p_4$, $k_1,\cdots, k_n$, $|{\cal M}_n^{PS}|^2$ (of order $\alpha^n$) is the
PS approximation to the
squared amplitude
for the decay $H\to 4\ell+n\gamma$, $\rmd\Phi_n$ is the exact phase
space for the decay and $\Pi(\{p\},\eps)$ is the Sudakov form factor
accounting for unresolved emission, {\it i.e.} soft (up to a cut-off energy $\epsilon$) and virtual corrections
 in the PS approximation. It is understood that the integral over the phase
space has a lower limit for the photon energies set to $\epsilon$, to
ensure the cancellation of the IR divergencies.

The quantities $\rmd\Phi_n$ and $\Pi(\{p\},\eps)$ read explicitly

\begin{equation}
\rmd{\Phi}_n =
\frac{1}{(2\pi)^{3n+8}}
\delta^{(4)}\left(P_H - \sum_{j=1}^4p_j - \sum_{i=1}^nk_i\right)
\prod_{j=1}^4\frac{\rmd^3{\vec{p}_j}}{2p_j^0}
\prod_{i=1}^n\frac{\rmd^3{\vec{k}_i}}{2k_i^0}
\label{eq:dPhi_n}
\end{equation}
\begin{equation}
  \Pi(\{p\},\eps) = \exp\left[{-\frac{\alpha}{2\pi} L I_\epsilon}\right]~~~~~~~~
  L\equiv\int\rmd\Omega_k(k^{0})^2\sum_{i,j=1}^4\eta_i\eta_j \frac{p_i\cdot
    p_j}{(p_i\cdot k)(p_j\cdot k)}
  \label{eq:SSF}
\end{equation}
In Eq.~(\ref{eq:SSF}), $L$ generates the soft/virtual collinear logarithms,
including also interferences effects of radiation coming from different
charged legs, and $I_\epsilon$, the integral of the Altarelli-Parisi
vertex for the branching $\ell\to\ell+\gamma$, generates the infrared
logarithms. It is explicitly given by:
\begin{eqnarray}
\!\!\!\!\!\! I_\epsilon \equiv \int_0^{1-\epsilon} dz \, \frac{1+z^2}{1-z} = 
 - 2 \ln \epsilon - \frac{3}{2} + 2\epsilon -\frac{1}{2}\epsilon^2 
\end{eqnarray}
In the definition of $L$, the integral is performed over
the angular variables of $k$, and $\eta_i$ equals $1$ if $i$ is an
anti-fermion or $-1$ if it is a fermion.

The integral over the phase space as in Eq.~(\ref{eq:dPhi_n}) is performed
after choosing a convenient set of independent variables and using
multi-channel MC importance sampling techniques to improve the
integration convergence and follow the peaking structure of the
partial decay width of Eq.~(\ref{eq:dGinftyPS}) to help events generation. The fully exclusive
information on final state particles momenta is kept. Details of the implementation
are given in appendix~\ref{app:phasespace}.

Before discussing the inclusion of NLO corrections
into Eq.~(\ref{eq:dGinftyPS}), it is interesting to point out that the
squared amplitudes with photon emissions are enhanced in regions of
the phase space where the photons are soft and/or collinear or where the
$Z$ propagators are resonating. In this perspective, a good
approximation to the exact matrix element can be written in the form\footnote{For the sake of simplicity, we 
consider the decay $H\to 2e2\mu +
n\gamma$, the generalization to $4e$ or $4\mu$ being straightforward.}:
\begin{eqnarray}
&&{\cal  M}_n^{\rm soft}(\{p\},\{k\},\{\sigma\},\{\tau\}) =\nonumber\\
&&c\sum_{\{{\cal P}\}}
\frac{J_{12}^\rho}{(p_1+p_2+Q_{\cal P})^2-m_Z^2}
\frac{J_{34,\rho}}{(p_3+p_4+R_{\cal P})^2-m_Z^2}
\prod_{i=1}^n
\frac{\eta_{{\cal P}_i}p_{{\cal P}_i}\!\cdot\!\varepsilon_{\tau_i}(k_i)}{p_{{\cal P}_i}\!\cdot\!k_i}
\label{eq:coherentsum}
\end{eqnarray}

In the previous equation, $c$ is a shorthand for the $HZZ$
coupling,
$\{\sigma,\tau\}$ label fermion and photon elicities,
$J^\mu_{ij}\equiv\bar u_{\sigma_i}(p_i)\gamma^\mu(g_V - g_A\gamma_5)v_{\sigma_j}(p_j)$,
$m_Z^2\equiv M_Z^2-i\Gamma_ZM_Z$ and $\varepsilon_\tau(k)$ are the photon
polarization vectors.
$\cal P$ is a $n$-dimensional vector whose $i^{th}$ component is the
index of the fermion to which the $i^{th}$ photon is attached and the
sum over $\cal P$ denotes all possible ways to share $n$ photons among
the four fermions. Finally,
$Q_{\cal P}$ is the sum of the momenta of the photons, for a given
$\cal P$, attached to the electron current ($R_{{\cal P}}$ to the
muon current).

Equation~(\ref{eq:coherentsum}) is derived from the amplitude for the
emission of photons in the soft limit but keeping the dependence on
the photon momenta  in the $Z$
propagators. The sum over the elicities of the squared
amplitudes of Eq.~(\ref{eq:coherentsum}) gives an approximation of the exact
squared matrix elements, coherently including also interferences among
diagrams. The final step to obtain $|{\cal M}_n^{PS}|^2$ of
Eq.~(\ref{eq:dGinftyPS}) from Eq.~(\ref{eq:coherentsum}) is to replace the
photon energy spectrum with the Altarelli-Parisi distribution for a better treatment of hard
collinear radiation.

Equation~(\ref{eq:dGinftyPS}), with the building blocks described above,
can then finally be improved to include exact NLO corrections
according to our master formula:
\begin{equation*}
  \rmd\Gamma_{\infty}^{\rm matched}
  = \frac{1}{2 M_H}\;F_{SV}\;\Pi(\{p\},\epsilon)
  \sum_{n=0}^\infty\frac{1}{n!}\;\left(\prod_{i=1}^n F_{H,i}\right)\;
  |{\cal M}_n^{PS}(\{p\},\{k\})|^2\rmd\Phi_n(\{p\},\{k\})
\end{equation*}
\begin{equation}
F_{SV} = 1+\frac{\rmd\Gamma^{NLO}_{SV}-\rmd\Gamma^{PS,\alpha}_{SV}}{\rmd\Gamma^{LO}}~~~~~~
F_{H,i} =
1+\frac{|{\cal M}_1^{NLO}(k_i)|^2 - |{\cal M}_{1}^{PS,\alpha}(k_i)|^2}{|{\cal M}_{1}^{PS,\alpha}(k_i)|^2}
\label{eq:masterformula}
\end{equation}

The correction factors $F_{SV}$ and $F_{H,i}$ carry the information of
the exact NLO calculation: $\rmd\Gamma_{SV}^{NLO}$ is the sum of the
virtual corrections of Eq.~(\ref{eq:hgv}) and Eq.~(\ref{eq:hw}) and soft real correction 
given by the first line of
Eq.~(\ref{eq:hg}), $\rmd\Gamma_{SV}^{PS,\alpha}$ is its PS approximation, 
{\it i.e.} the ${\cal O}(\alpha)$ term without any real hard photon of
Eq.~(\ref{eq:dGinftyPS}),  ${\cal M}_1^{NLO}(k_i)$ is the exact one-photon
bremsstrahlung amplitude and ${\cal M}_1^{PS,\alpha}(k_i)$ is its PS
approximation.

We want to remark that $F_{SV}$ and $F_{H,i}$ are by construction free
of collinear and/or infrared logarithms and that the
${\cal O}(\alpha)$ expansion of Eq.~(\ref{eq:masterformula}) exactly
coincides with the NLO calculation, without any double
counting. Furthermore, Eq.~(\ref{eq:masterformula}) is still fully
differential in the final state momenta and can be conveniently implemented in a MC
event generator.

Finally, we remark that the NLO virtual and real corrections used in
$F_{SV}$ and $F_{H,i}$ are strictly defined
only for $0$ or $1$ photon, while in Eq.~(\ref{eq:masterformula})  they
are used also when there are additional photons: this requires a
mapping of the $n$-photons phase space to $0$ or $1$ photon phase space. The
mapping is implemented in close analogy to the one described in
appendix A.2 of Ref.~\cite{Balossini:2006wc}, and here we do not
discuss it in further detail.

\section{Numerical results}
\label{sec:pheno}
In the present Section, we show and discuss the numerical results provided by 
our calculation, as obtained with the new tool {\tt Hto4l}. First, we show 
some tuned comparisons with the predictions of the reference code {\tt Prophecy4f}
at the level of NLO electroweak corrections. Then, we present our best 
predictions for various observables at NLOPS electroweak accuracy, 
as well as for Higgs production and decay in the presence of NLO QCD and electroweak
corrections matched to PS.

The results presented here are obtained using {\tt Prophecy4f v2.0}.\footnote{Available at
{\tt http://omnibus.uni-freiburg.de/\~{}sd565/programs/prophecy4f/prophecy4f.html}}
In both codes, we use the following set of input parameters

\begin{table}[ht]
\centering
\begin{tabular}{lll}
$\alpha(0) =1/137.03599911$ & 
$G_{\mu} = 1.16637~10^{-5}$ GeV$^{-2}$ &
$\mz=91.1876$~GeV \\
$\mw = 80.398$~GeV & 
$\gw = 2.141$~GeV &
$\gz = 2.4952$~GeV \\
$m_e=510.99892$~KeV &
$m_{\mu}=105.658369$~MeV &
$m_{\tau}=1.77684$~GeV \\
$m_u = 190$~MeV &
$m_c = 1.4$~GeV &
$m_{\rm top} = 172.5$~GeV \\
$m_d = 190$~MeV &
$m_s = 190$~MeV &
$m_b = 4.75$~GeV \\
\end{tabular}
\caption{\label{tab:input} Values of the input parameters used in the numerical calculations.}
\end{table}

The $M_{Z,W}$ and $\Gamma_{Z,W}$ are the running-width PDG values
which have to be converted to the fixed-width scheme adopted here
through, for example, the relations of Eq.~(7.2) of Ref.~\cite{Bredenstein:2006rh}.
As we work in the $G_\mu$ scheme, for the electromagnetic coupling constant 
we use the expression
\begin{eqnarray}
\alpha_{G_\mu} \, = \, \frac{\sqrt{2} \, G_\mu \, M_W^2 \sin^2\theta_W}{\pi} 
\end{eqnarray}
with $\sin^2\theta_W = 1 - M_W^2/M_Z^2$, in the calculation of 
the LO width and NLO weak corrections, while we use $\alpha(0)$ 
for the coupling of the photon to the external charged
particles.\footnote{This value is used for all the numerical results 
shown in the following, with the exception of the comparisons with {\tt Prophecy4f}, 
where we use $\alpha_{G_\mu}$ everywhere, to be consistent with the default 
choice of {\tt Prophecy4f}. } The top-quark width is set to the LO prediction in 
the SM, and a fixed width is employed in all the resonant propagators in the framework of
the complex mass scheme.

\subsection{NLO electroweak corrections: comparisons to {\tt Prophecy4f}}
\label{sec:nlo}

\begin{figure}[h]
\begin{center}
\includegraphics[width=0.7\textwidth]{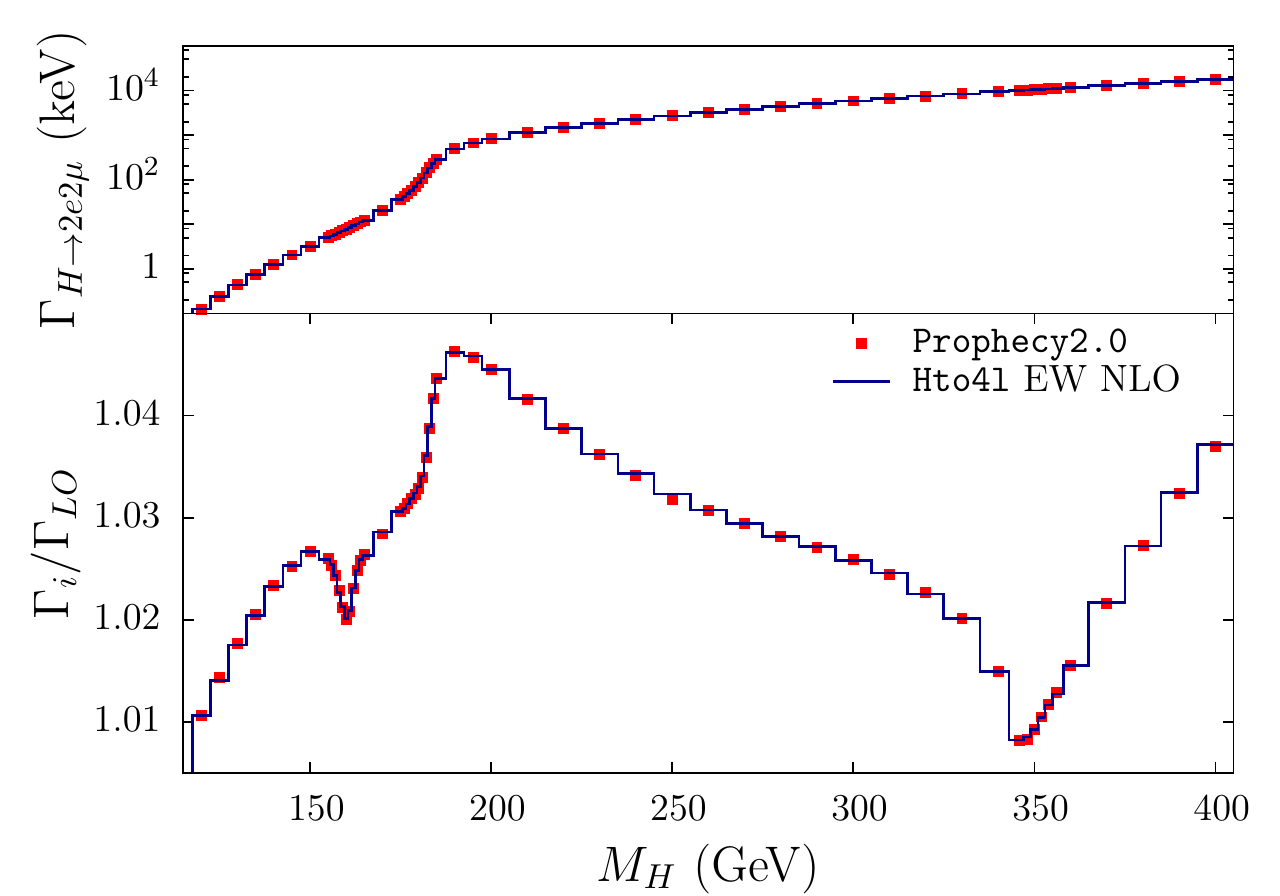}
\includegraphics[width=0.7\textwidth]{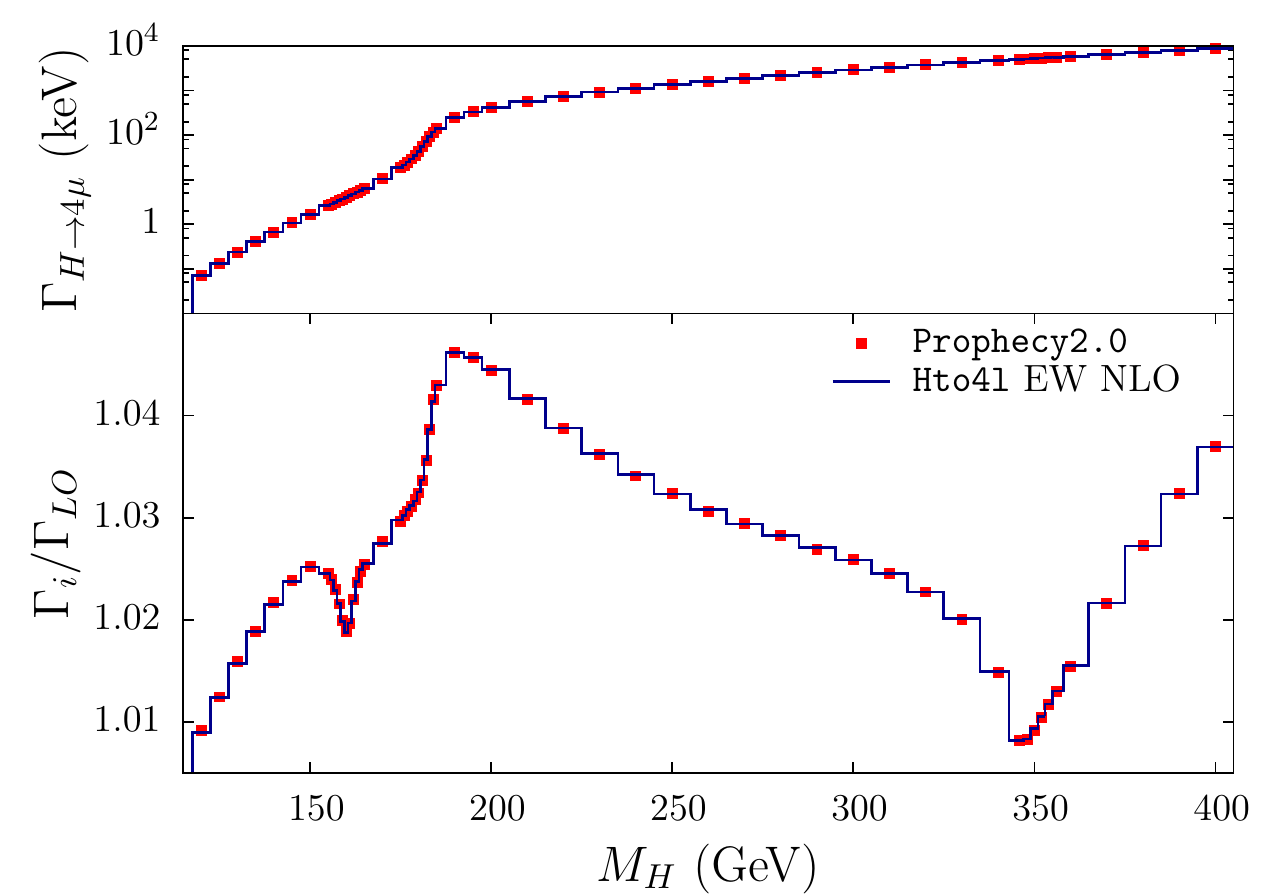}
\end{center}
\caption{\label{MassScan} Comparison between the NLO electroweak calculation of 
{\tt Prophecy4f} and {\tt Hto4l} for the decay widths $H \to 2e 2\mu$
(upper plot) and $H \to 4 \mu$ (lower plot), as a function of 
the Higgs mass in the range [125,~400]~GeV. For each plot, upper panel: absolute
predictions in KeV; lower panel: ratio between LO width and NLO EW
corrected  width.}
\end{figure}

\begin{figure}[t]
\begin{center}
\includegraphics[width=0.7\textwidth]{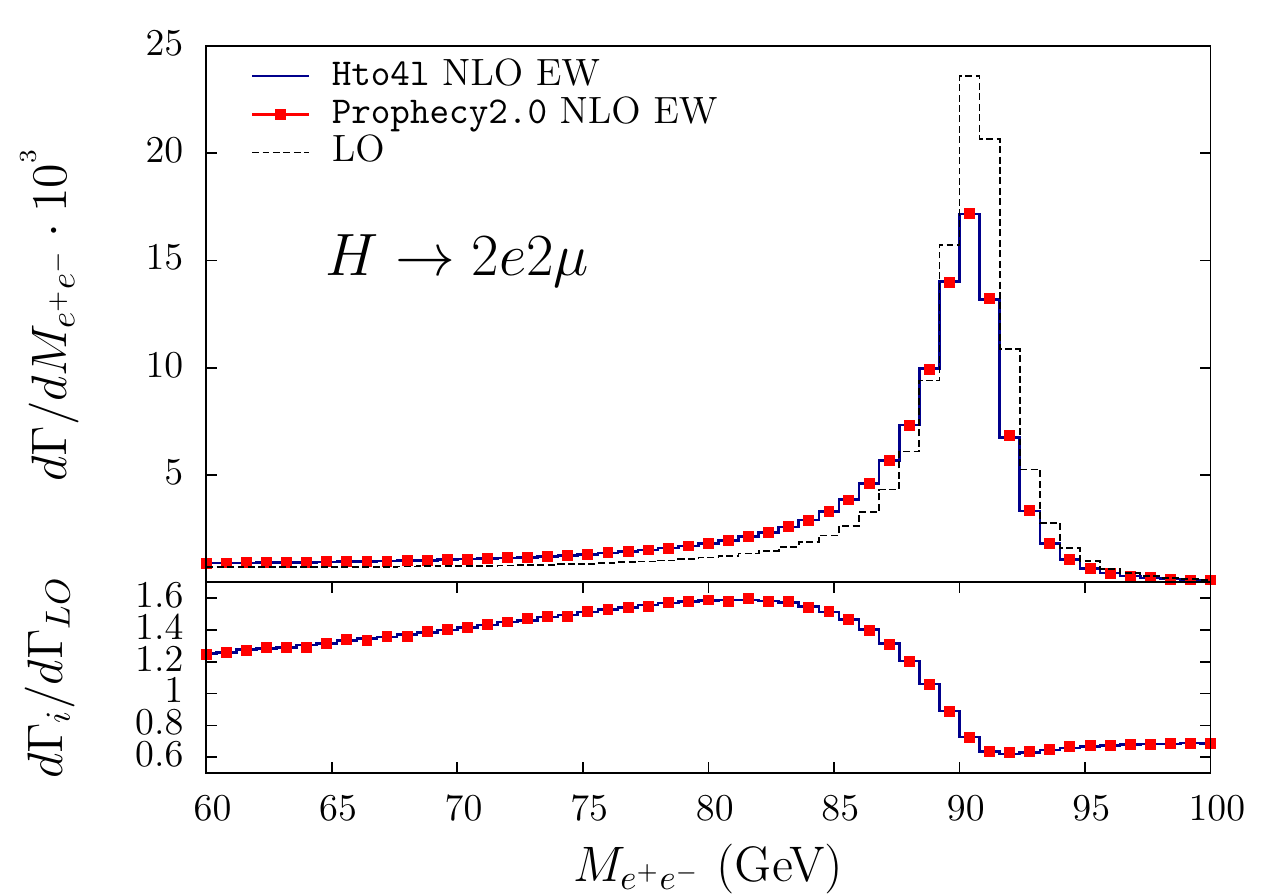}

\includegraphics[width=0.7\textwidth]{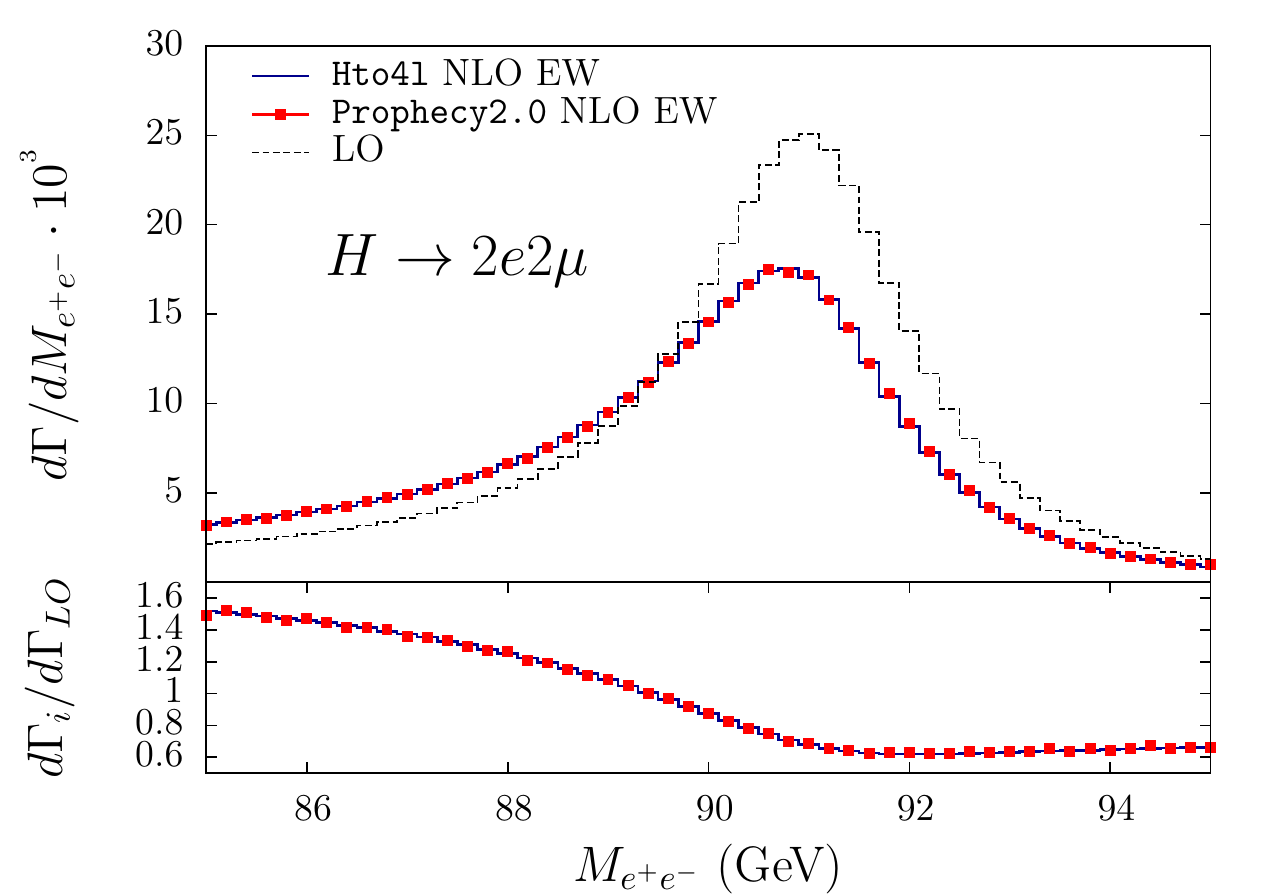}
\end{center}
\caption{\label{125m12eemm} Comparison between the NLO electroweak 
calculation of 
{\tt Prophecy4f} and {\tt Hto4l}
for the $e^+ e^-$ invariant mass (in the Higgs rest frame), 
in the range [60,~100]~GeV 
(upper plot) and in the range [85,~95]~GeV (lower plot). 
Predictions for the decay $H \to 2e 2\mu$ at $M_H = 125$~GeV.
Upper panels: absolute predictions for $d\Gamma / d M_{e^+ e^-}$; lower panels: relative effect 
of the NLO corrections.}
\end{figure}

\begin{figure}[ht]
\begin{center}
\includegraphics[width=0.7\textwidth]{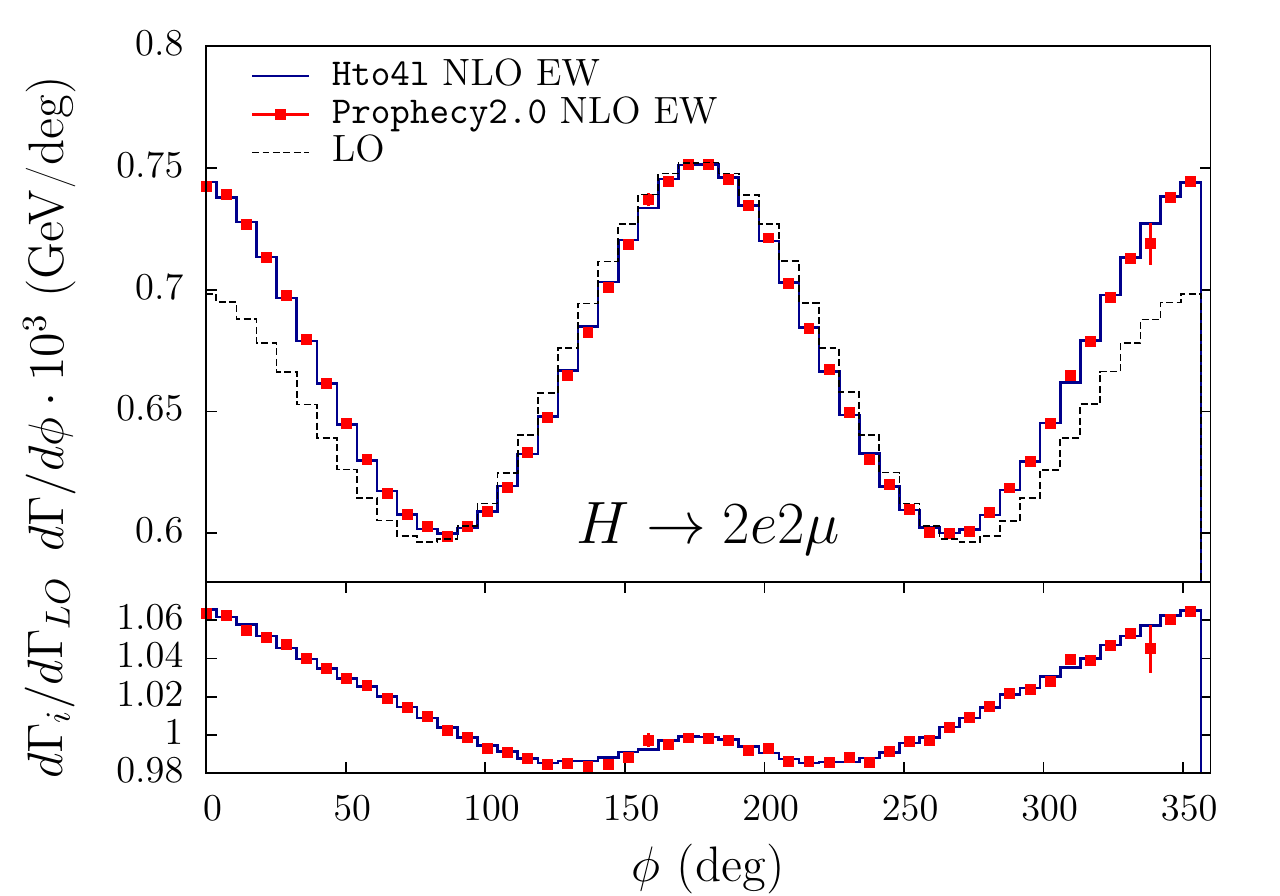}

\includegraphics[width=0.7\textwidth]{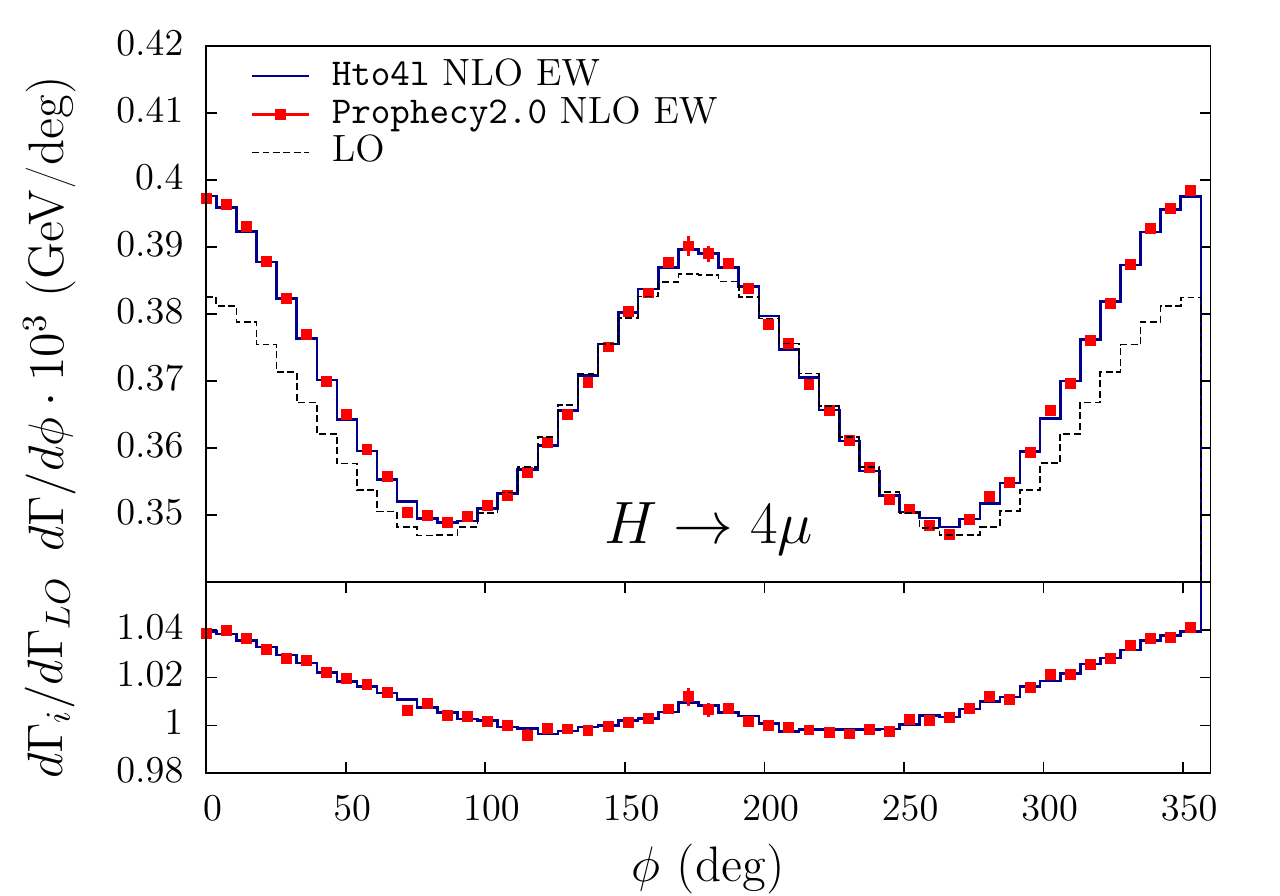}
\end{center}
\caption{\label{125phi1} Comparison between the NLO electroweak calculation 
of {\tt Prophecy4f} 
and {\tt Hto4l}
for the $\phi$ angle distribution  (in the Higgs rest frame) for the 
decay channels 
$H \to 2e 2\mu$ (upper plot) and $H \to 4 \mu$ (lower plot) at $M_H = 125$~GeV.
Upper panels: absolute predictions in GeV/deg; lower panels: relative effect 
of the NLO corrections.}
\end{figure}

A sample of the {\tt Prophecy4f} vs. {\tt Hto4l} comparisons 
at NLO electroweak accuracy is shown in Tab.~\ref{tab:NLOW} and 
in Figs.~\ref{MassScan}-\ref{125phi1}, in order to check the 
technical accuracy of our predictions in its different
aspects sketched in Sect.~\ref{sec:nloew}. Generally speaking, we 
observe very good agreement between our predictions and the
independent results of {\tt Prophecy4f}.

In Fig.~\ref{MassScan} we show the comparison for the NLO width in the 
leptonic decay channels $H \to 2e 2 \mu$ and $H \to 4 \mu$,\footnote{Analogous results are valid in the $H \to 4e$
  channel, which coincides for the integrated partial width with the
  $4\mu$ final state (apart from negligible mass effects).}
as a function of the Higgs mass 
in the range [125,~400]~GeV, together with the relative contribution due to the NLO 
electroweak corrections where the effect of mass thresholds 
present in the loop computation is particularly visible.
As can be seen, the two calculations perfectly agree. For the sake of clarity 
and completeness, 
we quote in Tab.~\ref{tab:NLOW} the predictions of the two codes 
for the decay channels $H \to 2 e 2 \mu$ and $H \to 4\mu$ for three specific 
values of the Higgs mass: the level of agreement is within the
statistical numerical uncertainty which is well below the 0.1\% accuracy.

\begin{table}[hbtp]
\centering
\begin{tabular}{| c || c | c |}
\hline
$M_H$/Final State & {\tt Prophecy4f} & {\tt Hto4l} \\
\hline
125~GeV/$H \to 2 e 2 \mu$ & 0.24151(8) & 0.24165(2) \\
\hline
140~GeV/$H \to  2 e 2 \mu$ & 1.2672(2) &  1.2667(1)\\
\hline
200~GeV/$H \to 2 e 2 \mu$ & 825.9(1) & 825.8(1) \\
\hline
125~GeV/$H \to 4\mu$ & 0.13324(2) & 0.13325(2) \\
\hline
140~GeV/$H \to 4\mu$ & 0.6713(1) & 0.6711(1) \\
\hline
200~GeV/$H \to 4\mu$ & 413.02(7) & 412.98(2) \\
\hline
\end{tabular}
\caption{\label{tab:NLOW}  Comparison between the NLO electroweak 
predictions of {\tt Prophecy4f} and our calculation ({\tt Hto4l}) for the 
Higgs decay width (in KeV), for different values of the Higgs mass and
final states. The numbers in parenthesis are the statistical uncertainty on 
the last digit
due to MC integration.}
\end{table}

In Fig.~\ref{125m12eemm} a comparison between 
 {\tt Prophecy4f} and {\tt Hto4l} is shown 
for the $e^+ e^-$ invariant mass (in the Higgs rest frame), in 
the range [60,~100]~GeV (upper plot) and in the range [85,~95]~GeV (lower plot).
The results refer to the decay channel $H \to 2e 2\mu$ for $M_H = 125$~GeV. 
Also in this case, the agreement between the two codes is remarkable, in spite of
the large effect due to the radiative corrections\footnote{For simplicity, in the present Section
we provide results for bare electrons only, {\it i.e.} in the absence of lepton-photon recombination effects.}. 
Actually, at and above the peak 
of the electron-pair invariant mass distribution the corrections are of the order of 
30\%, while for $M_{e^+ e^-}$ below $M_Z$ they can reach 50\%. The lowering of
the peak and the raising of a tail can be mainly ascribed to the photon radiation
off the leptons, as typical final-state radiation (FSR) effect observed around the
peak of 
resonant processes~\cite{CarloniCalame:2006zq,CarloniCalame:2007cd,Barze:2012tt,Barze':2013yca}.

A further comparison is given in Fig.~\ref{125phi1} for the distribution of 
the angle between the decay planes of the virtual $Z$ bosons in the
$H$ rest frame for the channels $H \to 2e 2\mu$ (upper plot) and
$H\to 4 \mu$ (lower plot) for $M_H = 125$~GeV, which is the observable of
main interest for spin-parity assignment.
For the $\phi$ angle we use the definition
\begin{eqnarray}
&&\cos \phi = \frac{(\mathbf{k}_{12} \times \mathbf{k}_1)\cdot(\mathbf{k}_{12} \times \mathbf{k}_3)}
{|\mathbf{k}_{12} \times\mathbf{k}_1||\mathbf{k}_{12} \times\mathbf{k}_3|} \\
&&{\mathrm{sgn}} (\sin \phi) = {\mathrm{sgn}} \left\{\mathbf{k}_{12} \cdot   \left[   (\mathbf{k}_{12} \times\mathbf{k}_1) \times (\mathbf{k}_{12} \times\mathbf{k}_3) \right]  \right\}
\end{eqnarray}
where $\mathbf{k}_{12} = \mathbf{k}_1 + \mathbf{k}_2$ and $\mathbf{k}_1$, $\mathbf{k}_2$,
$\mathbf{k}_3$, $\mathbf{k}_4$ are the three-momenta of the final-state
leptons.

Again the predictions of the two codes nicely agree.  The contribution
of the NLO corrections is particularly visible 
at the edges of the distribution, where it can reach the 5\% level for both the decay channels.

\subsection{Predictions at NLOPS electroweak accuracy}
\label{sec:nlops}

\begin{figure}[ht]
\begin{center}
\includegraphics[width=0.7\textwidth]{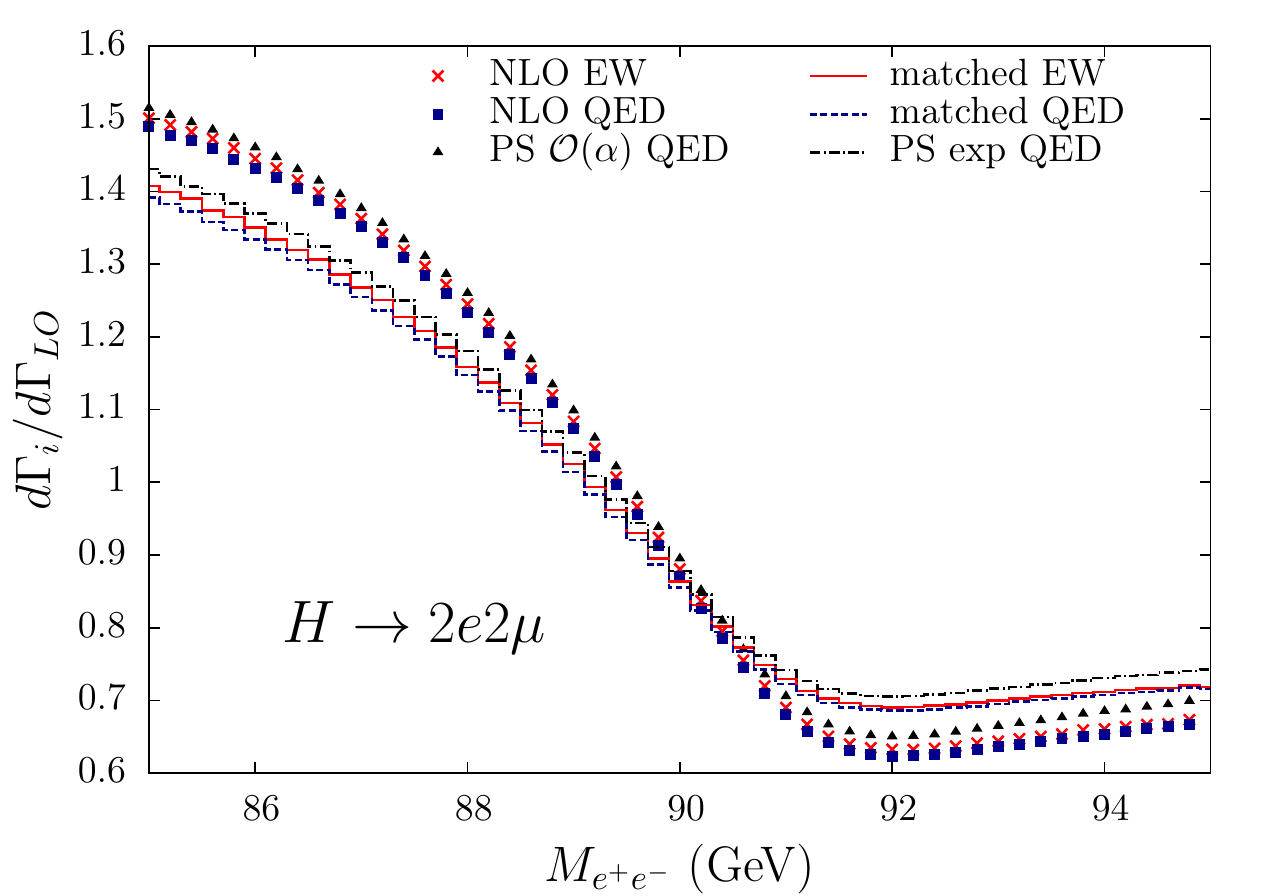}

\includegraphics[width=0.7\textwidth]{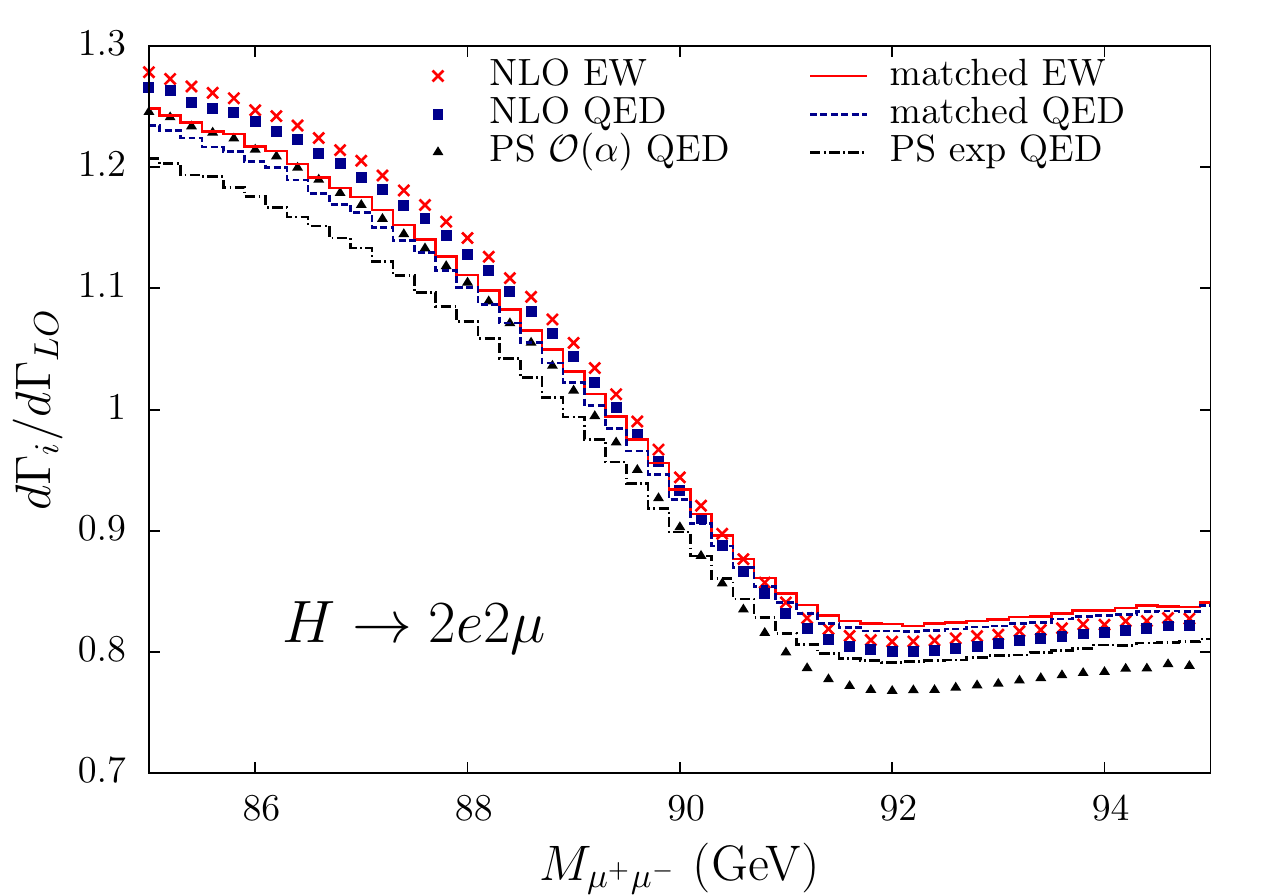}
\end{center}
\caption{\label{125m1234eemm} 
Relative contribution of the QED/electroweak corrections 
to the $e^+ e^-$ (upper plot) and $\mu^+ \mu^-$ (lower plot) 
invariant mass in the Higgs rest frame, 
 in the range [85,~95]~GeV. 
Predictions for the decay $H \to 2e 2\mu$ at $M_H = 125$~GeV.
The theoretical approximations corresponding to the different lines are explained in the text.}
\end{figure}

\begin{figure}[hbtp]
\begin{center}
\includegraphics[width=0.7\textwidth]{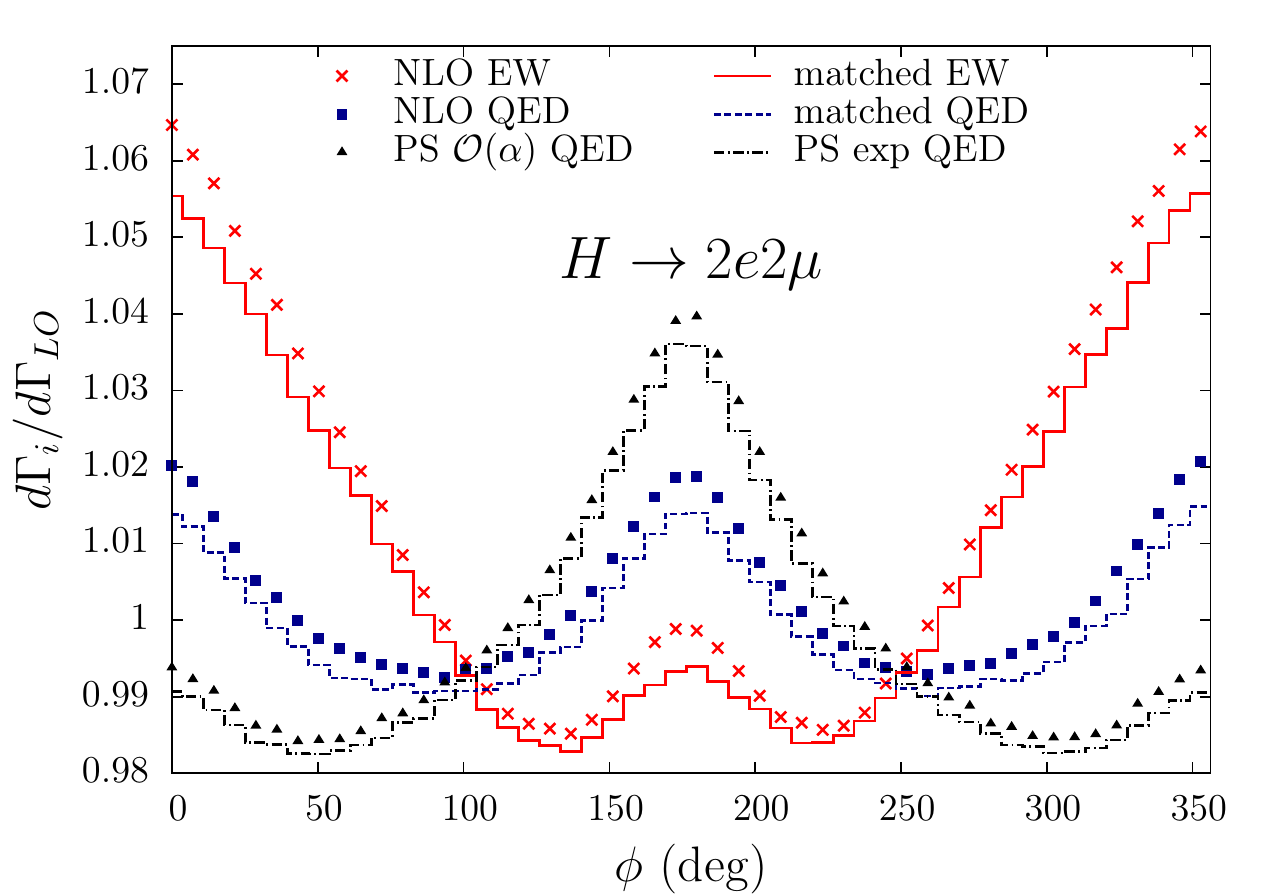}

\includegraphics[width=0.7\textwidth]{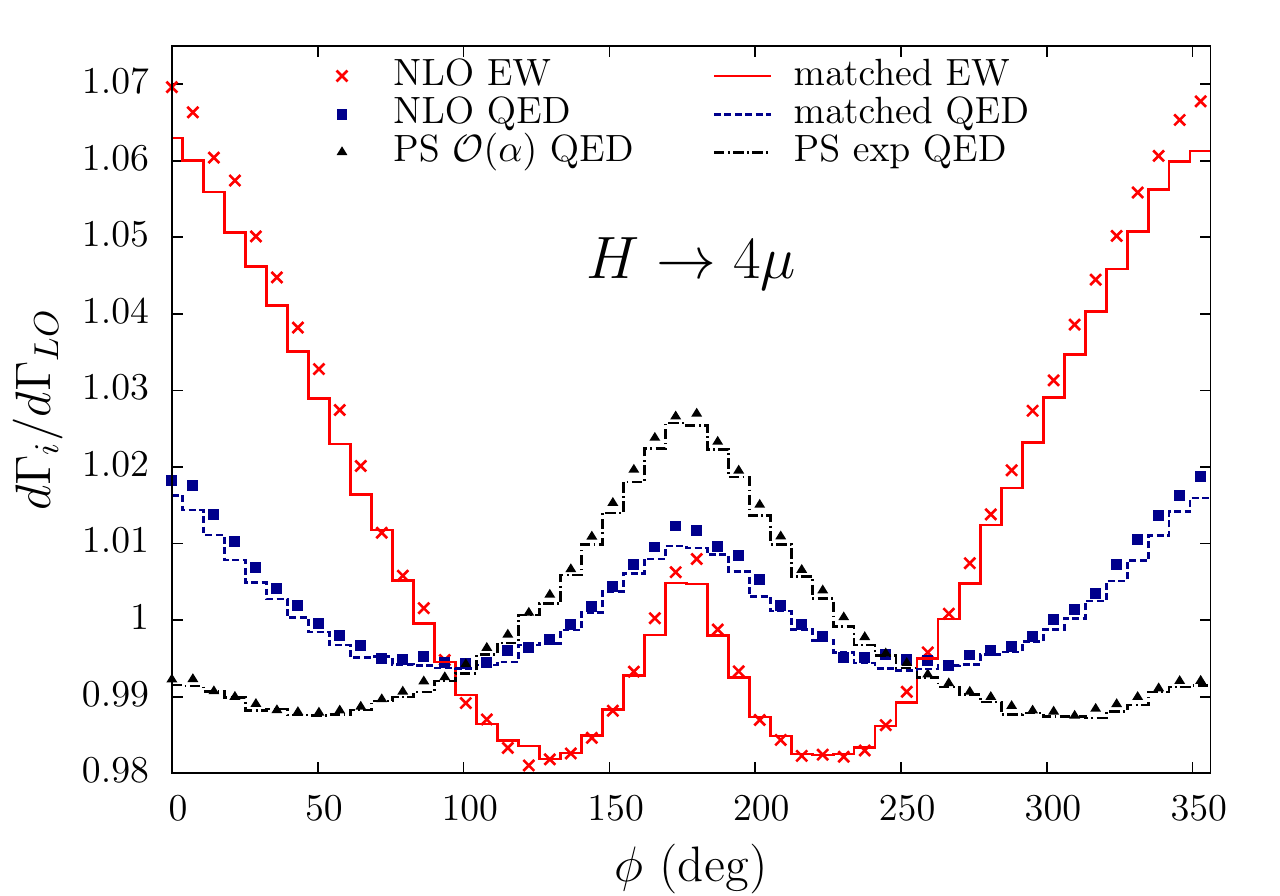}
\end{center}
\caption{\label{125phi1eff} The same as Fig. \ref{125m1234eemm} for the 
$\phi$ angle distribution in the decay channels $H \to 2e 2\mu$ (upper plot) and $H \to 4 \mu$ (lower plot) at $M_H = 125$~GeV.}
\end{figure}

\begin{figure}[t]
\begin{center}
\includegraphics[width=0.7\textwidth]{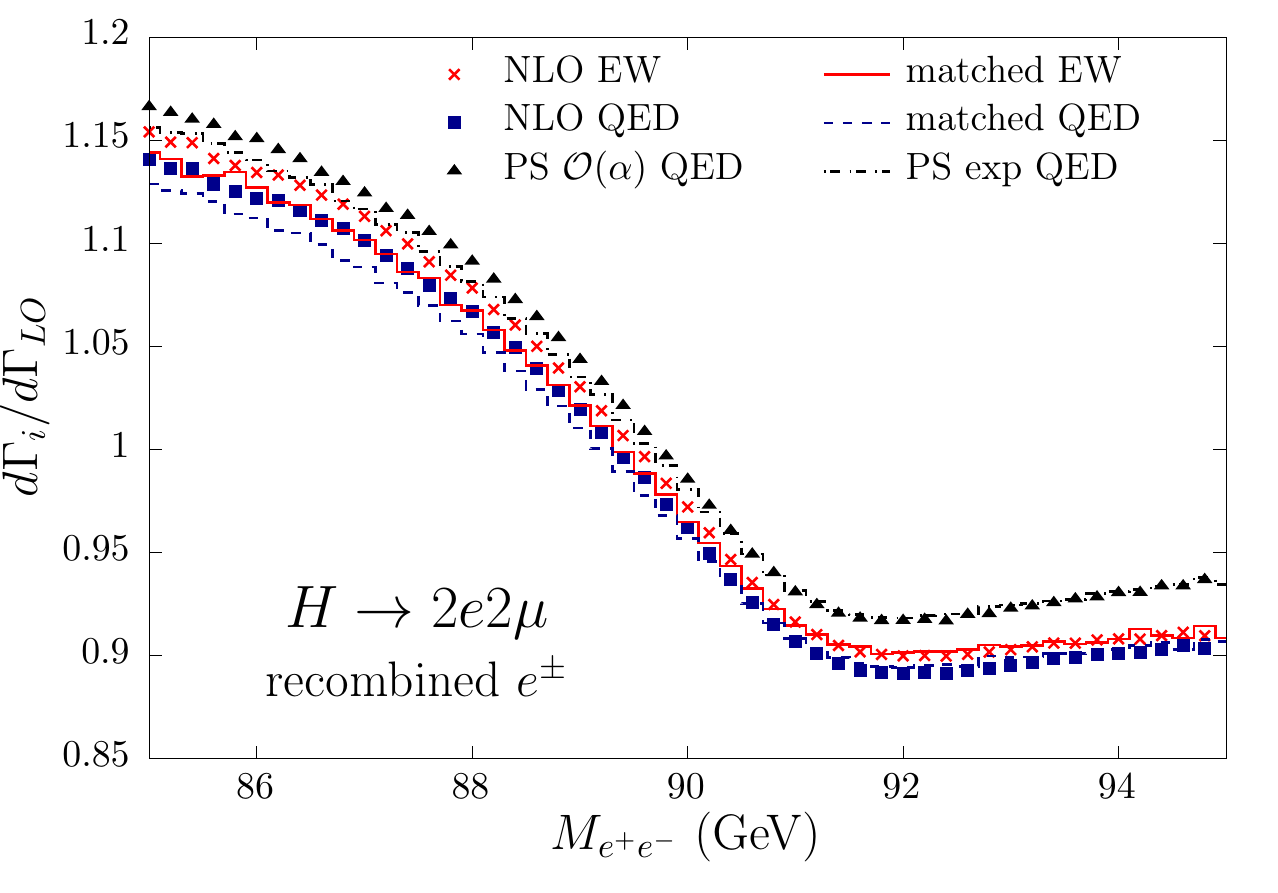}

\includegraphics[width=0.7\textwidth]{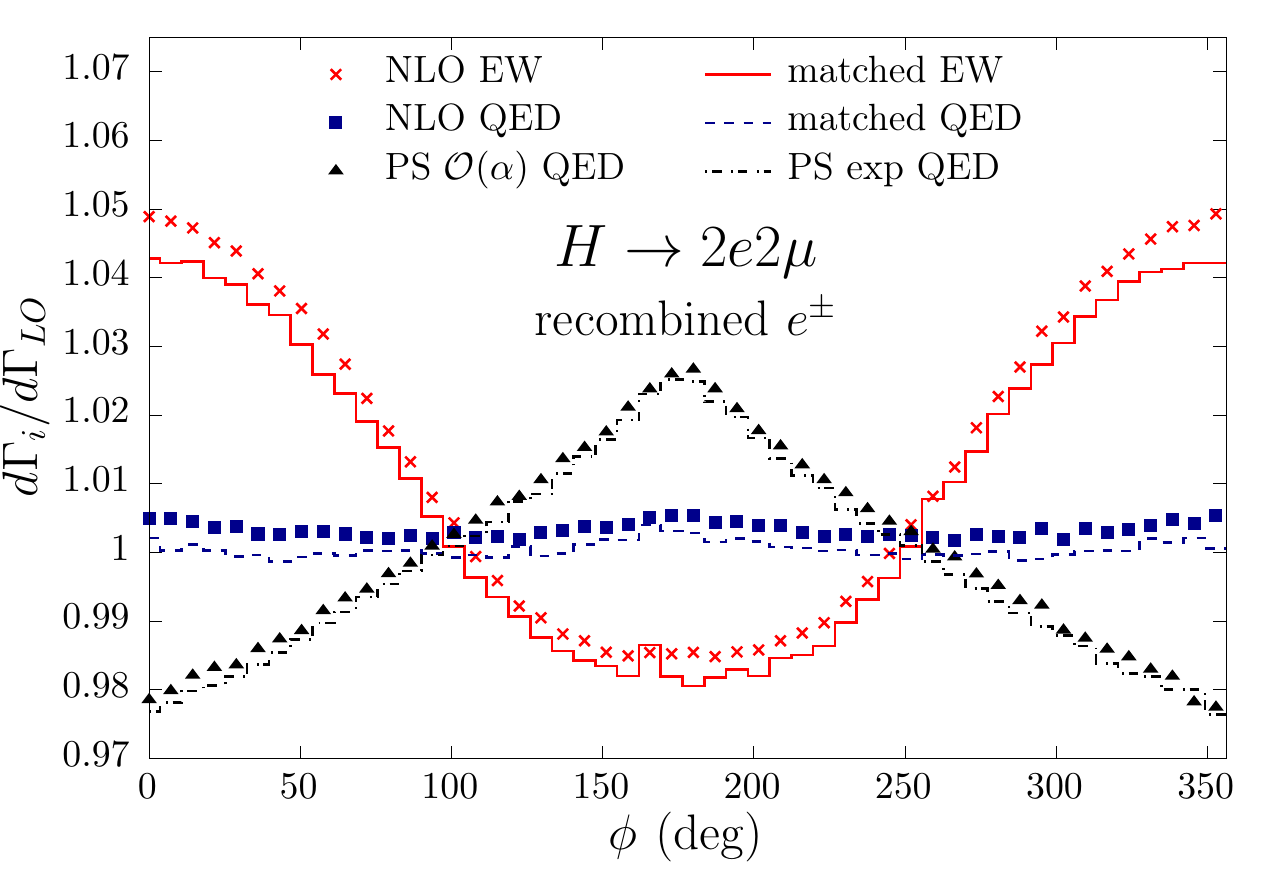}
\end{center}
\caption{\label{125m1234eemmrec} 
Relative contribution of the QED/electroweak corrections 
to the $e^+ e^-$ invariant mass (upper plot) and the $\phi$ angle distribution (lower plot) 
for recombined electrons and positrons. Predictions for the decay $H \to 2e 2\mu$ at $M_H = 125$~GeV in the Higgs rest frame.
The theoretical approximations corresponding to the different lines are explained in the text.}
\end{figure}

Some illustrative results obtained according to a number of variants of the theoretical 
approach described in Sect.~\ref{sec:mnlops} are given in Figs. \ref{125m1234eemm}-\ref{125m1234eemmrec}.
 In order to disentangle the impact of the different sources of correction, we consider the 
 results obtained according to the following levels of accuracy:
 \begin{enumerate}

\item the pure PS approximation for the decay width as in Eq.~(\ref{eq:dGinftyPS}), associated to 
multiple photon emission in the soft/collinear limit;
 
\item the ${\cal O} (\alpha)$ truncated approximation of Eq.~(\ref{eq:dGinftyPS}), describing one photon 
radiation in the PS framework;

\item the complete NLO electroweak calculation; 
 
\item the NLO QED calculation, given by the gauge-invariant subset of electromagnetic contributions 
within the full set of electroweak corrections;
 
\item the NLO electroweak corrections matched to the QED PS, as in Eq.~(\ref{eq:masterformula});

\item the NLO QED corrections matched to the QED PS, {\it i.e.} the QED gauge-invariant realization
of Eq.~(\ref{eq:masterformula}).
 
 \end{enumerate}

The comparison between approximations 1. and 2. is useful to quantify the 
higher-orders contribution due to photon emission beyond 
${\cal O} (\alpha)$, while the difference between options 3. and 4. is
a measure of pure weak loop corrections, the difference between
approximations 2. and 3. is an estimate of non-logarithmic
${\cal O}(\alpha)$ QED terms plus pure weak loop corrections. The 
comparison between approximations 3. and 5., as well as between 4. and 6., allows us to check that the NLOPS 
matching procedure correctly preserves the effect of QED exponentiation as given by the difference between 
options 1. and 2. Moreover, the results of 1. vs. those of 5. 
and of 3. vs. those of 5. provide an estimate of the accuracy
of the predictions available in the literature for Higgs physics at the LHC, in particular of 
 of the process-independent, widely used code
{\tt PHOTOS}~\cite{Barberio:1993qi}, which describes multiple photon 
emission but does not include exact NLO electroweak corrections, and
of {\tt Prophecy4f}, that does not take into account the contribution of exclusive QED exponentiation.

In Fig. \ref{125m1234eemm} we show the relative contribution of the different theoretical 
approximations discussed above for the $e^+ e^-$ (upper plot) and $\mu^+ \mu^-$ (lower plot)
invariant mass in the Higgs rest frame, 
 in the range [85,~95]~GeV. The results refer to the process $H \to 2e 2\mu$ for $M_H = 125$~GeV, 
 according to a bare lepton definition. By inspection of Fig. \ref{125m1234eemm} we can draw
 the following conclusions: the NLO corrections to the lepton invariant masses are quite large, since 
 they amount to about 50\% (30\%) to the $e^+ e^-$ ($\mu^+ \mu^-$) invariant mass below the peak 
 and about 30\% (20\%) at and above it. They are largely dominated by the enhanced leading logarithmic
 contributions of QED nature $\propto \alpha \log (M_Z^2/m_\ell^2)$, as can be inferred from the 
 comparison between the results of the pure ${\cal O} (\alpha)$ PS algorithm and those of the NLO 
 QED/electroweak calculations. From this comparison, one can also conclude that the ${\cal O} (\alpha)$ 
 non-logarithmic QED 
 terms contribute at the some per cent level, both for the $e^+ e^-$ and $\mu^+ \mu^-$ 
 invariant mass, whereas the pure weak loops have a much smaller effect, not exceeding the
 1\% level.
 
 The large impact of NLO QED corrections, which significantly modify the shape of the invariant mass distribution, 
 translates in a relevant contribution due to higher-order photonic corrections. Multiple photon emission 
 is of the order of 10\% for the $e^+ e^-$ final-state and at the level of some per cents for the $\mu^+ \mu^-$
 case, as a consequence of the different magnitude of the lepton-photon collinear logarithm. 
 It can also be noticed that QED exponentiation reduces the impact of NLO corrections 
 and that the NLOPS matching correctly preserves the size 
 of multiple photon emission.
 
Quite different conclusions derive from the analysis of Fig. \ref{125phi1eff}, which shows 
the relative corrections of the different theoretical recipes on the $\phi$ angle distribution 
for the $H \to 2e 2\mu$ and $H \to 4\mu$ decays. 
For such an observable, the pure ${\cal O} (\alpha)$ PS approximation significantly 
underestimates the contribution of NLO EW
corrections for $\phi$ close to $0^{\degree}$ and $360^{\degree}$, while it provides 
an overestimate around $180^{\degree}$. Actually, it can be noticed that the $\phi$ angle
distribution receives a non-negligible contribution from fixed-order non-logarithmic terms and that, more
importantly, is particularly sensitive to pure weak corrections, which set the correct overall size and 
shape of the radiative corrections. On the other hand, the effect of QED exponentiation is moderate, 
varying between a few per mille to about 1\%.

For completeness, we show in Fig.~\ref{125m1234eemmrec} results for the invariant mass of 
the $e^+ e^-$ pair and the $\phi$ angle distribution (for the process $H \to 2e 2\mu$) under the 
more realistic experimental condition of calorimetric or recombined 
electrons and positrons. In this case, we replace the three-momentum of the $e^\pm$ with the effective 
momentum $\bf{p} = \bf{p_{e^\pm}} + \bf{p_{\gamma}}$ for each photon satisfying 
the condition $\Delta R_{e^\pm \gamma} = \left( \Delta\eta_{e^\pm \gamma}^2
+   \Delta\phi_{e^\pm \gamma}^2 \right)^{1/2} \leq 0.1$, as typically done by LHC experiments, 
where $\Delta\phi_{e^\pm \gamma}$ is the lepton-photon separation angle in the 
transverse plane. As can be seen from Fig.~\ref{125m1234eemmrec} in
comparison to Fig. \ref{125m1234eemm} and Fig. \ref{125phi1eff}, 
 the contribution of the radiative corrections is largely reduced, as expected, when 
switching from bare to recombined electrons/positrons. For the $e^+ e^-$ invariant mass, the 
corrections are reduced by about a factor of three, almost independently of the 
considered theoretical approximation, and preserve their shape. However, 
non-negligible corrections still remain under the calorimetric condition, of about +15\% 
in the left tail of the invariant mass and of the order of -10\% around the peak of 
the distribution, when considering the most accurate matched predictions. 
In comparison to the case of 
bare electrons, the effect of QED exponentiation for dressed electrons 
reduces to about 1\% in the tail and at the per mille
level at and above the peak.

More interestingly, 
the QED relative corrections to the $\phi$ angle distribution are substantially modified 
both in size and shape by the recombination effects, whereas the full electroweak 
predictions receive a slight size reduction and a less pronounced shape modification. 
In particular, we checked
through detailed numerical inspections that the especially visible difference in shape 
between the pure PS and the diagrammatic QED predictions is of virtual origin and has to be ascribed 
to the QED pentagons, which are exactly included in the Feynman
diagram calculation and only (crudely) approximated in the soft/collinear limit in
the PS calculation. To some extent, we expect that the rich angular correlations
introduced by pentagon diagrams is only poorly reproduced by the PS approximation. 
All in all, the results shown in 
the lower plot of Fig.~\ref{125m1234eemmrec} 
reinforce the already noticed particularly relevant r\^ole played by loop contributions 
with complex topology, both of QED and weak nature, to obtain reliable predictions for the $\phi$ angle 
observable.

To summarize, the main conclusion of this Section is that both NLO electroweak and higher-order QED corrections, as well as 
their combination, are relevant for reliable simulations of the most important observables
considered in precision studies of the Higgs sector at the LHC.

\subsection{Interface to {\tt POWHEG}: results for production and decay}
\label{sec:pwhg}

\begin{figure}
\begin{center}
  \includegraphics[width=0.7\textwidth]{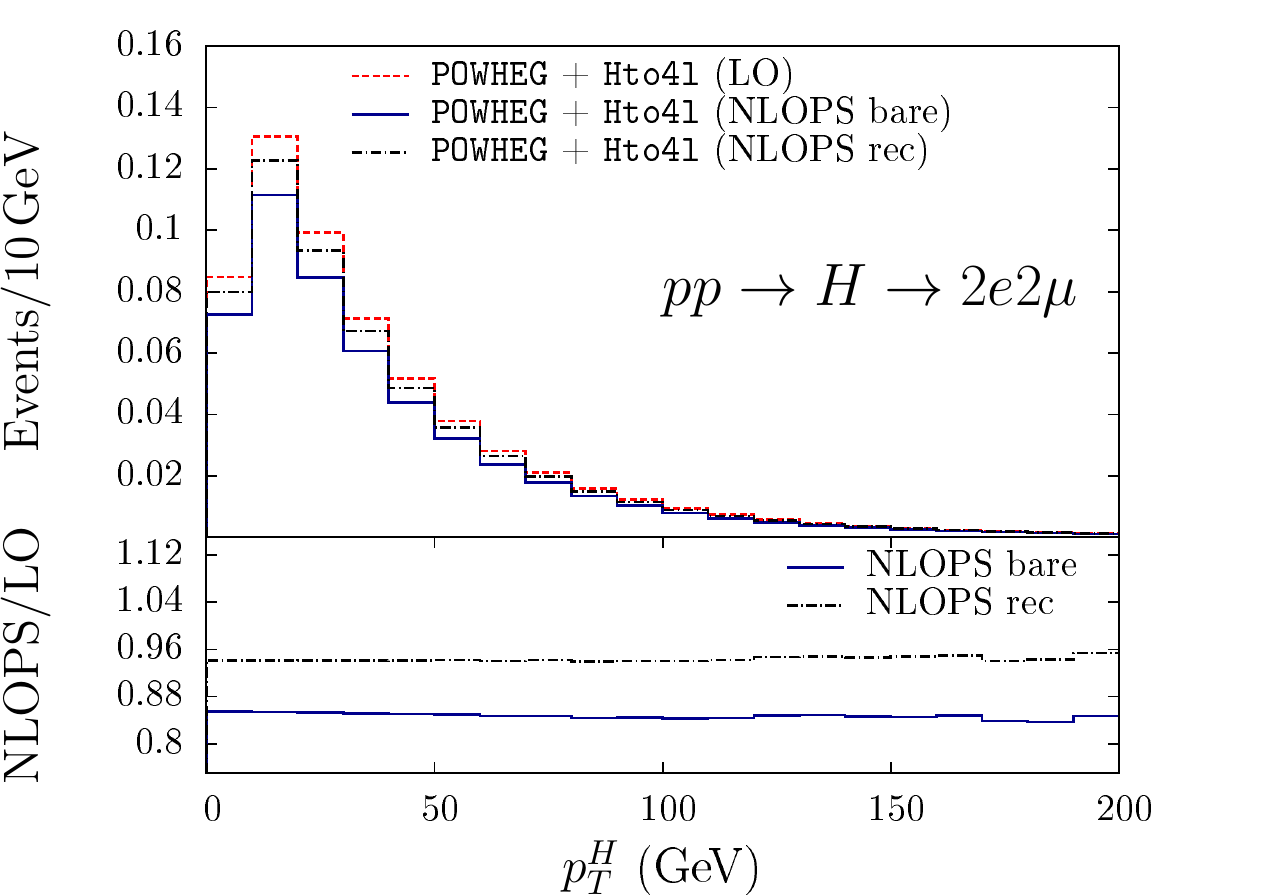} 

  \includegraphics[width=0.7\textwidth]{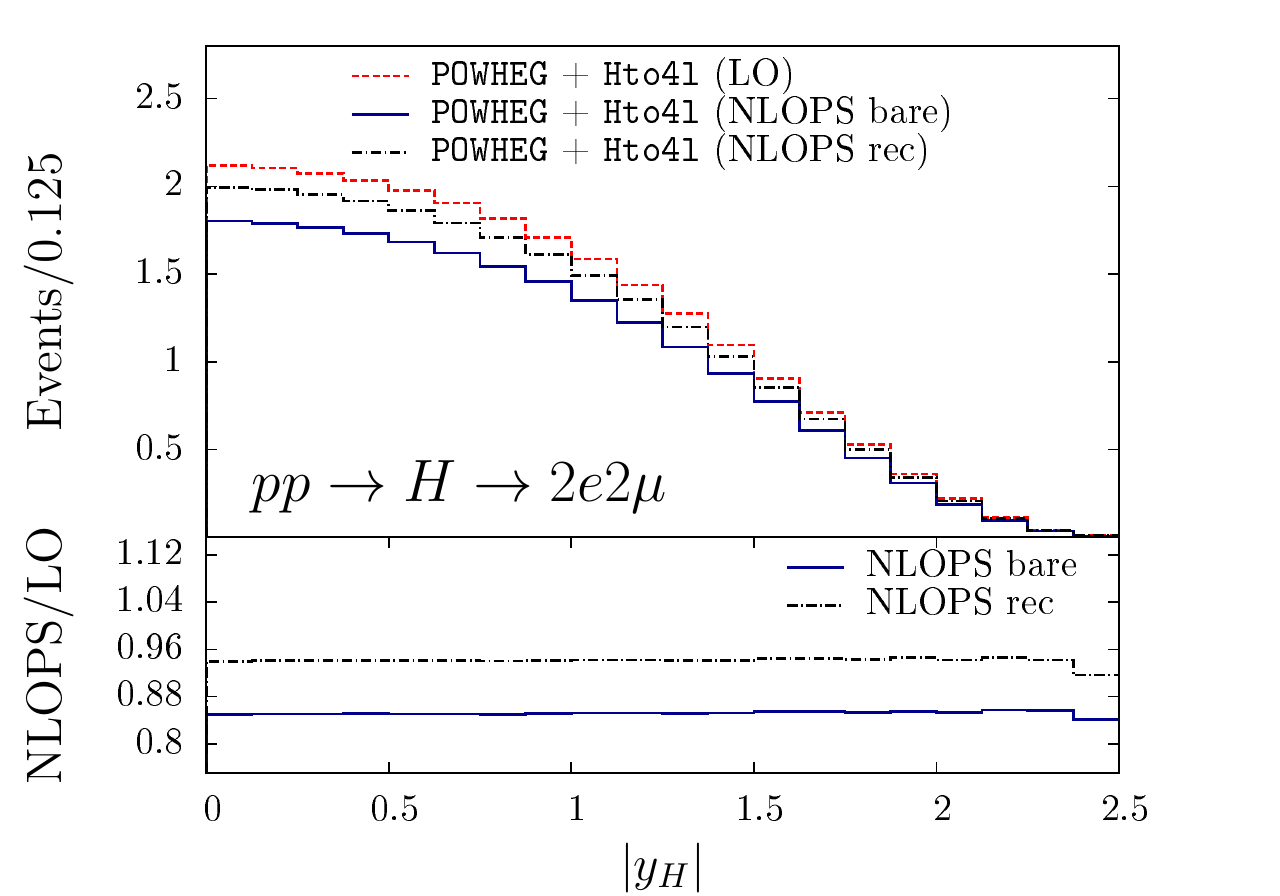}
\end{center}
\caption{\label{pthyh}  Comparison between the results obtained with {\tt POWHEG} + {\tt Hto4l} (Born) + {\tt PYTHIA $\!\!$v6} 
(red dashed line) 
and  {\tt POWHEG} + {\tt Hto4l} (NLOPS) + {\tt PYTHIA $\!\!$v6} (blue solid 
and black dash-dotted lines) for the transverse momentum (upper plot) and rapidity (lower plot) of the Higgs boson. 
In the lower panels the relative contribution of NLOPS electroweak corrections 
for bare (solid line) and recombined (dash-dotted line) leptons is shown.}
\end{figure}

\begin{figure}
\begin{center}
  \includegraphics[width=0.7\textwidth]{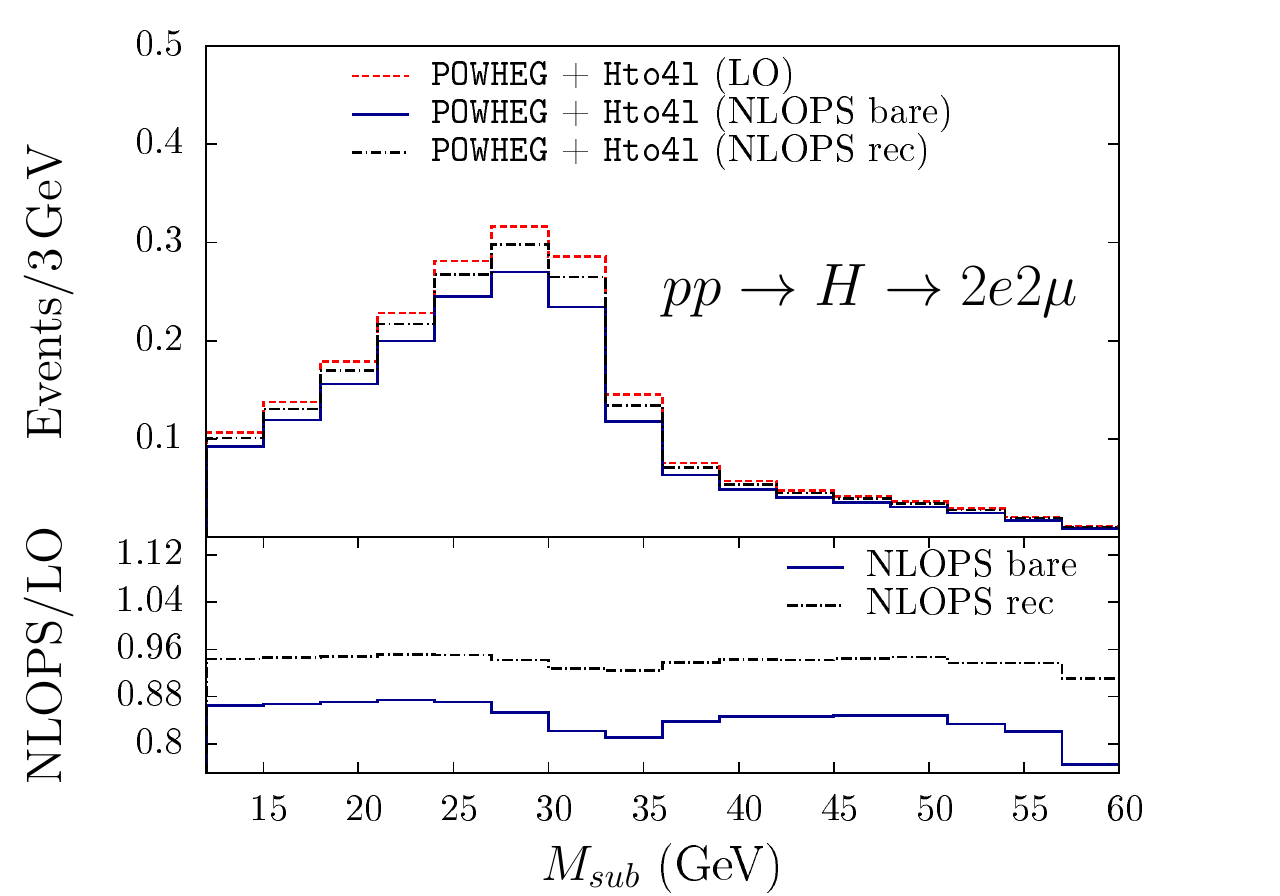}

  \includegraphics[width=0.7\textwidth]{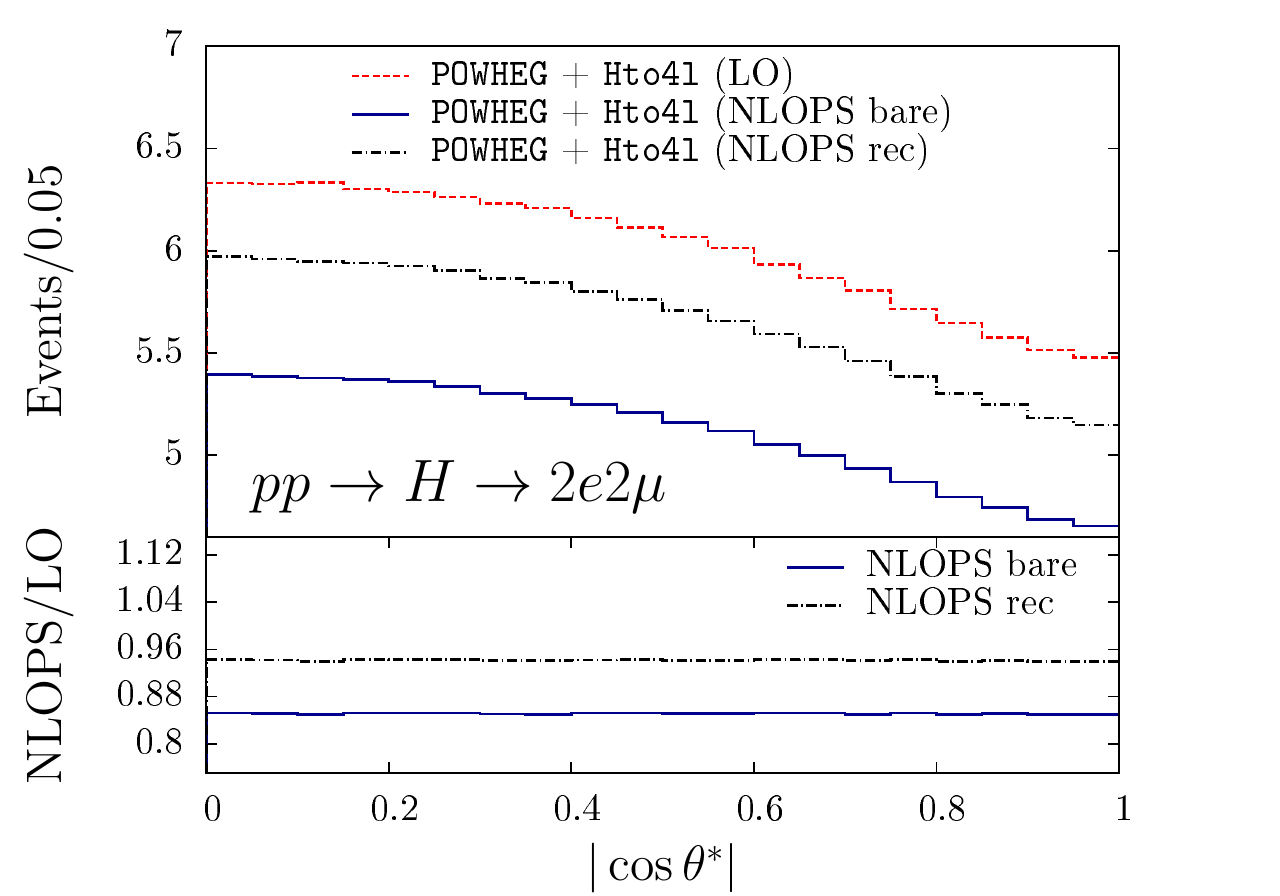}
\end{center}
\caption{\label{msubcos} The same as in Fig.~\ref{pthyh} for the invariant mass of the subleading lepton pair 
(upper plot) and the cosine of the angle of the leading lepton pair in the four-lepton rest frame with
respect to the beam axis (lower plot).}
\end{figure}

\begin{figure}
\begin{center}
  \includegraphics[width=0.7\textwidth]{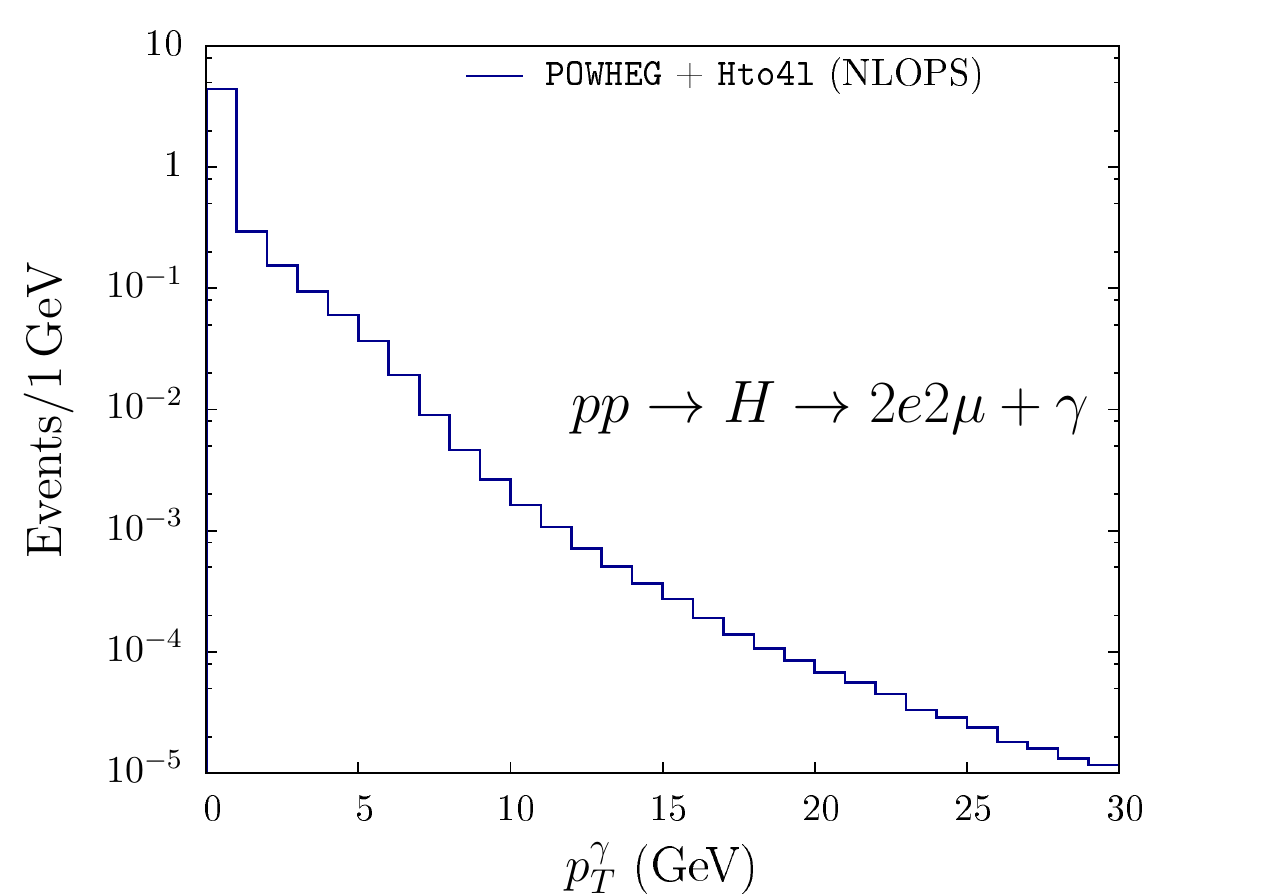}

  \includegraphics[width=0.7\textwidth]{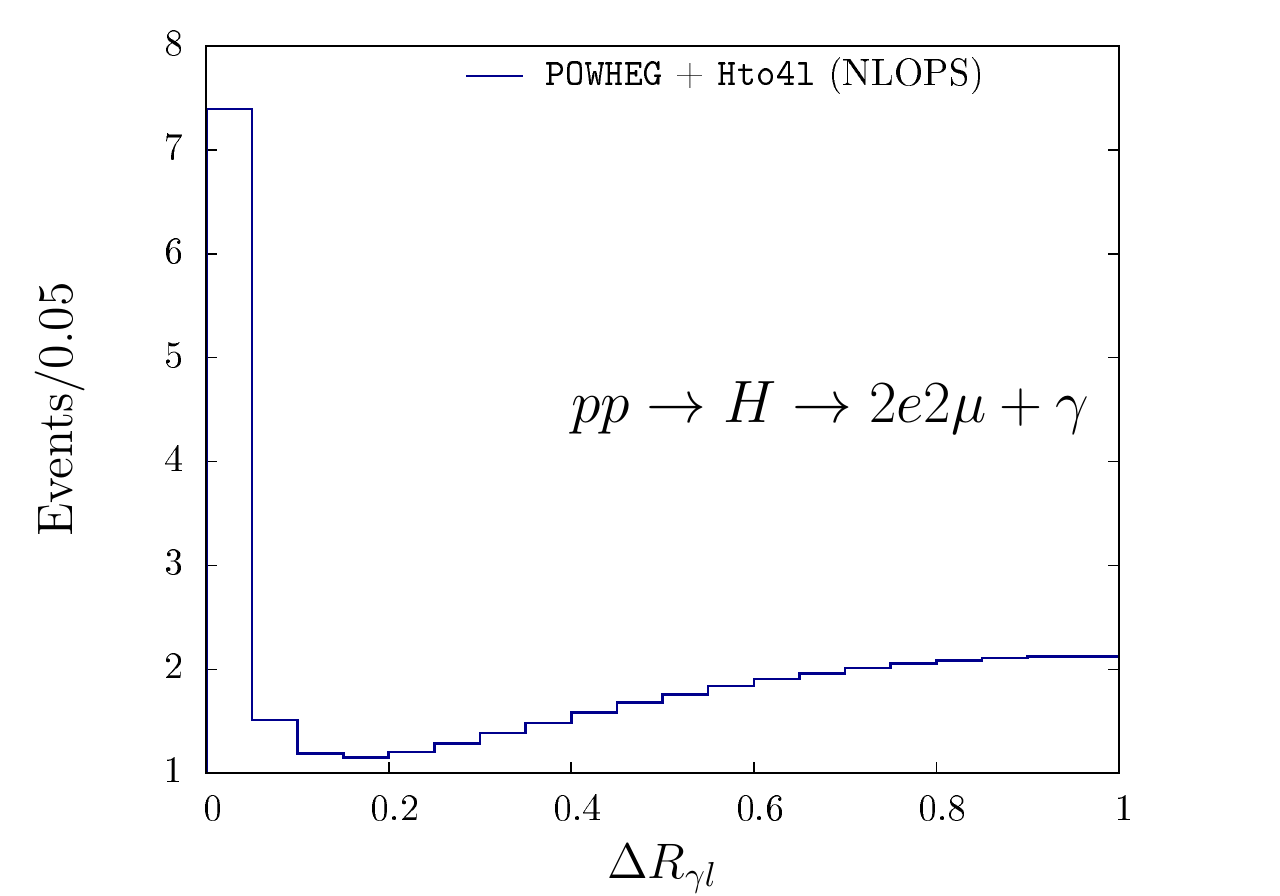}
\end{center}
\caption{\label{photons} Distribution of the transverse momentum of
  the hardest photon (upper plot) and the angular separation 
between the hardest photon and the closest lepton (lower plot) obtained using
{\tt POWHEG} + {\tt Hto4l} (NLOPS) + {\tt PYTHIA $\!\!$v6}. 
The minimum photon energy is $E_{\gamma}^{\rm min} = 1$~GeV.}
\end{figure}

In order to facilitate phenomenological studies of Higgs boson production and 
decay in the presence of both QCD and electroweak contributions, 
we have implemented an interface which allows to use our code in association with any event 
generator describing Higgs production. In Figs.~\ref{pthyh}-\ref{photons} 
we show a sample of illustrative results obtained 
by interfacing {\tt Hto4l} with {\tt POWHEG}~\cite{Frixione:2007vw} for the simulation of Higgs boson 
production in gluon-gluon fusion. We use the {\tt POWHEG} 
version with NLOPS accuracy in QCD~\cite{Alioli:2008tz} from the {\tt POWHEG BOX} 
framework~\cite{Alioli:2010xd} and we consider 
Higgs production in proton-proton collisions at a c.m. energy of 8~TeV\footnote{
However, as we are interested to study the relative 
impact of electroweak corrections dominated by contributions of 
the kind $\alpha^n \log^n (M_Z^2/m_\ell^2)$, the results shown in 
the following are in practice independent of the c.m. energy.}.
The events generated by  {\tt POWHEG} are interfaced to {\tt Hto4l}
according to the following procedure: 
\begin{itemize}
\item generate unweighted events for the process
  $pp \to H (+j)$  in the Les Houches format using {\tt POWHEG}, 
where $H$ is an on-shell Higgs boson and $j$ stands for the extra parton of the NLO QCD calculation;
	\item the Les Houches file is read event by event by {\tt Hto4l} and the particles momenta 
	         are stored in the generic common block structure introduced in Ref.~\cite{Boos:2001cv}; 
\item each event is decayed into the selected channel in the $H$ rest
  frame, using {\tt Hto4l}. After boosting the decay products back to
  the laboratory frame, the events including production and decay are written in a file in the Les Houches format.
\end{itemize}

The Les Houches file can be finally passed to a shower event generator for QCD 
showering and hadronization. In our examples we use {\tt PYTHIA $\!\!$v6.4}~\cite{Sjostrand:2006za} as QCD PS. 
According to the above procedure, the $pp \to H \to 4\ell$ process is 
treated in narrow width approximation, as it is the case for 
a 125~GeV Higgs boson, and factorized in on-shell Higgs production and decay.

In our analysis we consider, for definiteness, the decay channel $H
\to 2e 2\mu$ and the following observables: 
the transverse momentum $p_{{T}}^{H}$ and rapidity $y_{H}$ of the Higgs boson (Fig.~\ref{pthyh}), 
the invariant mass of the subleading lepton pairs and the magnitude of the cosine of the decay
angle of the leading lepton pair in the four-lepton rest frame with respect to the beam
axis $|\cos\theta^*|$ (Fig.~\ref{msubcos}). The leading pair is
defined as the lepton pair with invariant 
mass closest to the $Z$ boson mass and its angle is obtained by summing 
the three-momenta of the two leptons. For the {\tt POWHEG} calculation 
of Higgs production in gluon fusion, we use the PDF set {\tt MSTW2008nlo68cl}~\cite{Martin:2009iq} 
with factorization/renormalization scale $\mu_R = \mu_F = M_H$.
The values of the other input parameters are the same as the ones given in Tab.~\ref{tab:input}.
The results shown in the following refer to a sample of $1,2\cdot
10^8$ unweighted events and to the same selection cuts adopted in
Ref.~\cite{Aad:2014tca}
and correspond to bare (solid line) and recombined (black dash-dotted line) leptons. In the
latter case, we recombine photons with both electrons and muons, 
in analogy to the selection criteria adopted in the experimental study of 
Ref.~\cite{Aad:2014tca}, if the condition $\Delta R_{\ell \gamma} \leq 0.1$ is satisfied.

In Fig.~\ref{pthyh} and Fig.~\ref{msubcos} we show the comparison between 
the predictions obtained using {\tt POWHEG} interfaced to our code at LO and NLOPS electroweak 
accuracy. It can be noticed that the contribution due to NLOPS electroweak corrections is 
almost flat and of about $-15 \, (-5)$\% for $p_{{\rm T}}^{H}$, $y_{H}$ and $|\cos\theta^*|$ when 
considering bare (recombined) leptons, while 
the invariant mass of the subleading lepton pairs receives a varying correction of size 
between $-20 \, (-10)$\% and $-10 \, (-5)$\% for bare (calorimetric) leptons, respectively.

In Fig.~\ref{photons} we show the 
results for two observables which are fully exclusive over QED
radiation and which can be easily treated 
in our approach. The results correspond to the process
$p p \to H \to 2e2\mu + n\,\gamma$, with $E_{\gamma}^{\rm min} = 1$~GeV, for
which we show the transverse momentum of the hardest photon 
and the angular separation between the hardest photon and the closest lepton, that
exhibit the expected features of photon emission in radiative events.

\section{Conclusions}
\label{sec:conc}

In this work we have presented a precision calculation of the SM Higgs boson decay
into four charged leptons, in view of improved measurements of the properties of the
Higgs particle at the LHC Run II. Our approach is based on the computation of the
full one-loop electroweak corrections supplemented with the contribution of multiple photon
emission taken into account according to a fully exclusive QED PS algorithm. 
Our results, which have a NLOPS electroweak accuracy, are available in the 
form of a new event generator, {\tt Hto4l}, that can be easily interfaced to any 
QCD program simulating Higgs production.

We have cross-checked our NLO electroweak corrections against the predictions 
of the reference code {\tt Prophecy4f} and found perfect agreement. We have also
shown that both NLO electroweak and higher-order QED corrections, as well as
their interplay, are necessary for actually precise simulations of the variety of 
observables involved in Higgs physics at the LHC. This provides the main 
novel theoretical feature of our work, which goes beyond the presently available 
results  limited to the fixed-order approximation or to a leading logarithmic QED 
modeling. The second relevant aspect is given by the possibility of interfacing 
{\tt Hto4l} to any generator describing Higgs boson production, thus allowing 
simulations of Higgs production and decay in the presence of higher-order 
QCD and electroweak corrections. In this respect, we have shown some 
illustrative results obtained in terms of the combined usage of {\tt POWHEG} 
and {\tt Hto4l}.

Our results can find application in precision measurements of the Higgs 
boson mass, spin-parity determination and tests of the SM at the level of
differential cross sections in the future run of the LHC. They can be 
generalized to other processes yielding four leptons in 
hadronic collisions, like {\it e.g.} $ p p \to H \to W^{(*)}W^{(*)} \to 2\ell 2\nu$ or
$p p \to Z^{(*)}Z^{(*)} \to 4 \ell$.

\acknowledgments

This work was supported in part by the Research Executive Agency (REA) 
of the European Union 
under the Grant Agreement number PITN-GA-2010- 264564 (LHCPhenoNet), and by 
the Italian 
Ministry of University and Research 
under the PRIN project 2010YJ2NYW. 
F.P. wishes to thank the CERN PH-TH Department for partial support and 
hospitality during several stages of the work. The authors thank the 
Galileo Galilei Institute for theoretical physics and INFN for partial
support during the completion of this research.

\appendix
\section{Phase space parameterisation and integration}
\label{app:phasespace}

The $4+n$ bodies phase space as in Eq.~(\ref{eq:dPhi_n}) is integrated
according to
standard multi-channel MC techniques, combined with importance sampling
to reduce the variance of the integral and help event
generation\footnote{Here we consider only the decay $H\to 2e2\mu$, the
generalization to 4 identical leptons being straightforward.}. The
first step is to generate a photon multiplicity $n$ and associate
$n_1$ ($n_2$) photons to the electron (muon) current ($n_1+n_2=n$), defining
the channel of the multi-channel integration. The phase space is then conveniently
split into two decaying objects to follow the $Z$ propagators, namely
\begin{eqnarray}
&&\rmd\Phi(P_H;p_1,\cdots,p_4,k_1,\cdots,k_n) = (2\pi)^6\rmd Q^2_{Z_1}\rmd
  Q^2_{Z_2}\rmd\Phi(P_H;P_{Z_1},P_{Z_2})\times
  \nonumber\\
&&~~~~\rmd\Phi(P_{Z_1};p_1,p_2,k_{1},\cdots,k_{n_1})\;\rmd\Phi(P_{Z_2};p_3,p_4,k_{n_1+1},\cdots,k_{n_1+n_2})
\label{eq:phz1z2}
\end{eqnarray}
where $P_{Z_i}$ ($P_{Z_i}^2=Q_{Z_i}^2$) are the momenta of the virtual $Z$
bosons.

We refrain from writing explicitly the simple $1\to 2$ decay phase
spaces of Eq.~(\ref{eq:phz1z2}) and we focus instead on the case where at least
one photon is present. As discussed in appendix A.3 of
Ref.~\cite{Balossini:2006wc}, an efficient sampling of photons collinear to
final state leptons is a non trivial task, because the directions of
the leptons are known only after all the momenta are generated. In
Ref.~\cite{Balossini:2006wc} we adopted a solution based on a properly
chosen multi-channel strategy. Here we adopt a different and elegant solution, which
consists in writing the phase space in the frame where the leptons are
back-to-back, {\it i.e.} $\vec{p_a}=-\vec{p_b}$ (see for example~\cite{Schonherr:2008av,JadachLect,Jadach:1999vf}).

Omitting overall numerical factors for brevity, the building block we are interested in is
\begin{eqnarray*}
\rmd\Phi(P;p_a,p_b,k_1,\cdots,k_r) &=& \delta^{(4)}\left(P - p_a-p_b - \sum_{i=1}^rk_i\right)
\frac{\rmd^3{\vec{p}_a}}{p_a^0}\frac{\rmd^3{\vec{p}_b}}{p_b^0}
\prod_{i=1}^r\frac{\rmd^3{\vec{k}_i}}{k_i^0}\\
&\equiv& \delta^{(4)}(P - Q - K)\,\delta\Phi
\end{eqnarray*}
where we defined $Q=p_a+p_b$, $K=\sum_{i=1}^rk_i$ and $\delta\Phi$
contains the infinitesimal phase space element divided by the final
state particle energies. It is usually
understood that all the variables are expressed in the frame where $P$
is at rest, but we want to express them where $Q$ is at rest. In order
to do that, the previous equation
can be further manipulated by inserting the following identities
\begin{eqnarray}
  &&\rmd^4Q\,\delta^{(4)}(Q-p_a-p_b)=1\nonumber\\
  &&\rmd s^\prime\,\delta(Q^2-s^\prime) = 1\nonumber\\
&&\rmd^4 P\,\delta^{(3)}(\vec{P})\,\delta(P^0-\sqrt{s})=2\sqrt{s}\;\rmd^4 P\,\delta^{(3)}(\vec{P})\,\delta(P^2-s)=1
\label{eq:unities}
\end{eqnarray}
which help to make explicit the Lorentz invariance of the phase space
element.

With the help of Eq.~(\ref{eq:unities}) and appropriately rearranging
the terms, we can write
\begin{eqnarray}
  &&\rmd\Phi(P;p_a,p_b,k_1,\cdots,k_r) = \delta\Phi\;
  \rmd^3\vec{Q}\delta^{(4)}(Q-p_a-p_b)\delta^{(3)}(\vec{P})\nonumber\\
  &&2\sqrt{s}\;\rmd^4P\delta^{(4)}(P-Q-K)\delta(P^2-s)\rmd s^\prime\rmd Q^0\delta(Q^2-s^\prime)=\nonumber\\
  &&=\delta\Phi\sqrt{\frac{s}{s^\prime}}\;\delta^{(3)}(\vec{P})\delta^{(4)}(Q-p_a-p_b)\delta((Q+K)^2-s)\rmd^3\vec{Q}\rmd
  s^\prime=\nonumber\\
&&=\rmd s^\prime\frac{s}{s^\prime} \delta^{(4)}(Q-p_a -
  p_b)\delta((Q+K)^2-s) \frac{\rmd^3{\vec{p}_a}}{p_a^0}\frac{\rmd^3{\vec{p}_b}}{p_b^0}
  \prod_{i=1}^r\frac{\rmd^3{\vec{k}_i}}{k_i^0}=\nonumber\\
&&=
  \frac{s^\prime}{2s}\beta_a\rmd\Omega_a\frac{1}{1+\frac{\sum_{i=1}^rk^0_i}{\sqrt{s^\prime}}}\prod_{i=1}^r\frac{\rmd^3{\vec{k}_i}}{k_i^0}
\label{eq:deltas}
\end{eqnarray}
In the cascade of identities~(\ref{eq:deltas}) we used the result
$\rmd^3\vec{Q}\;\delta^{(3)}(\vec{P}) = (s^\prime/s)^{\frac{3}{2}}$
(see~\cite{JadachLect}) and we made use of Lorentz invariance. In the last
identity it is understood that all the variables are expressed in the
frame where $Q=p_a+p_b$ is at rest and $s^\prime=Q^2$, $s=P^2$, $\beta_a$
is the speed of particle $a$ and
$\rmd\Omega_a=\rmd\cos\theta_a\rmd\phi_a$. The big advantage of the
last equation is that the lepton momenta $p_a$ and $p_b$ lie on the
same direction defined by $\cos\theta_a$ and $\phi_a$, hence all
photons can be generated along this direction to sample the collinear
singularities. Once all particle momenta are generated, they can
be boosted back to the rest frame of the decaying Higgs boson.

One last remark concerns the integration limits of the phase space. As
mentioned in Sect.~\ref{sec:mnlops}, photon energies should be
generated larger than the infrared cut-off $\epsilon$ in the Higgs
frame, which is a non Lorentz invariant cut. Since the minimum photon
energy can not be determined {\it a priori} in the frame where $Q$ is
at rest (because $Q$ itself depends on the photons momenta), we decide
to generate photon energies starting from $0$ to cover the whole phase
space and then, once boosted back, cut the
event if a photon enegy falls below~$\epsilon$. Finally, in order to flatten
the infrared divergence, we choose to sample the photon energies according
to the function
$$f(\omega)\propto \begin{cases} \frac{1}{\omega}
  ~~~~\omega\ge\epsilon^\prime\\ \frac{1}{\epsilon^\prime} ~~~~\omega<\epsilon^\prime  \end{cases}$$
where $\epsilon^\prime$ is a guessed (and tuned for efficiency)
minimum energy.

\bibliographystyle{JHEP}
\bibliography{Hto4l-rev}

\end{document}